\documentclass[usenatbib,useAMS]{mn2e}

\usepackage{amsfonts}
\usepackage{amsmath}
\usepackage{wasysym}
\usepackage{graphicx}
\usepackage{longtable}
\usepackage{lscape}

\begin{document}

\label{firstpage}

\title[Interpolations of the IGIMF]{Simple interpolation functions for the galaxy-wide stellar initial mass function and its effects in early-type galaxies}

\author[Dabringhausen]{
J. Dabringhausen$^{1}$ \thanks{E-mail: joerg@sirrah.troja.mff.cuni.cz}\\
$^{1}$ Charles University, Faculty of Mathematics and Physics, Astronomical Institute, V  Hole\v{s}ovi\v{c}k\'ach 2,\\
 CZ-180 00 Praha 8, Czech Republic}

\pagerange{\pageref{firstpage}--\pageref{lastpage}} \pubyear{2019}

\maketitle

\begin{abstract}
The galaxy-wide stellar initial mass function (IGIMF) of a galaxy is thought to depend on its star formation rate (SFR). Using a catalogue of observational properties of early-type galaxies (ETGs) and a relation that correlates the formation timescales of ETGs with their stellar masses, the dependencies of the IGIMF on the SFR are translated into dependencies on more intuitive parameters like present-day luminosities in different passbands. It is found that up to a luminosity of approximately $10^9 \ {\rm L}_{\odot}$ (quite independent of the considered passband), the total masses of the stellar populations of ETGs are slightly lower than expected from the canonical stellar initial mass function. However, the actual mass of the stellar populations of the most luminous ETGs may be up to two times higher than expected from an SSP-model with the canonical IMF. The variation of the IGIMF with the mass of ETGs are presented here also as convenient functions of the luminosity in various passbands. 
\end{abstract}


\section[Introduction]{Introduction}
\label{sec:introduction}

One of the fundamental parameters in astronomy is the distribution function of stellar masses. The present-day stellar mass function (PDMF) of a galaxy mostly determines how much light a galaxy emits in which parts of the electromagnetic spectrum. Furtermore, the PDMF establishes how many stars of which mass evolve in a galaxy at a given time, and thereby the chemical composition of the matter reinserted into its interstellar medium. Thus, the PDMF of a given galaxy plays a crucial role for its observed parameters and its further evolution, and it varies from galaxy to galaxy, even though the characteristics of the PDMF correlate with galaxy type.

There are two two key aspects which determine the PDMF of a galaxy. The first aspect is the age spectrum of the stars, which is given by the star formation history (SFH). The SFH quantifies how many stars formed at which time, and thereby which of the stars that once formed in the galaxy have already evolved into remnants. The second aspect is the mass spectrum of of the stars that formed in the galaxy at any time, which is given by the integrated galaxy-wide stellar initial mass function (IGIMF). The IGIMF was introduced by \citet{Kroupa2003} as the overall stellar initial mass function in all star-forming regions of a galaxy within a characteristic time interval $\delta t$. Thus, varying SFHs and varying IGIMFs may both be the reason for the well-known large differences in the PDMFs of galaxies.

In practice, a variation of the SFH among galaxies is beyond dispute, and apparent by the fact that spiral and irregular galaxies show indications of ongoing star formation, while most elliptical galaxies of the same size do not.

A variation of the IGIMF is more controversial; in particular since it boils down to a variation of the stellar initial mass function (IMF). The IMF was first introduced by \citet{Salpeter1955}, and quantifies the mass spectrum of stars formed in a single event. The complete composition of the stellar population in a single event happens quite fast, and the result is a so-called embedded star cluster with mass $M_{\rm ecl}$. The IGIMF on the other hand is a composite population of many IMFs. It comprises the IMFs of all star clusters in a galaxy that formed over the time-span $\delta t$. In principle, the IMFs that make up the IGIMF can vary from star cluster to star cluster. 

Theoretically, the IMF is indeed expected to vary for various reasons. One possible reason is that the critical mass for a gas cloud to collapse under its own gravity depends on its temperature and its density \citep{Larson1998}. Another reason is that proto-stars may collide and merge more often in dense environments \citep{Murray1996}. However, observationally, there was for a long time no firm evidence for a variation of the IMF, so that a universal IMF of the form
\begin{equation}
\xi_{\rm IMF} (m) =k \, k_i \, m^{-\alpha_i},
\label{eq:IMF}
\end{equation}
with 
\begin{align}
\nonumber \alpha_1 = 1.3  & {\rm \ if \ } 0.1 \le  \frac{m}{\rm{M}_{\odot}} < 0.5,\\
\nonumber \alpha_2 = 2.3  & {\rm \ if \ } 0.5 \le  \frac{m}{\rm{M}_{\odot}} < 1.0,\\
\nonumber \alpha_3 = 2.3  & {\rm \ if \ } 1.0 \le  \frac{m}{\rm{M}_{\odot}} \le m_{\rm max},
\end{align}
was suggested for all star formation events \citep{Kroupa2001, Kroupa2013}. In this equation, $m$ is the initial stellar mass, $m_{\rm max}$ is the maximum initial stellar mass, the factors $k_i$ ensure that the IMF is continuous where the power changes and $k$ is a normalization constant. The normalisation constant is set such that the integral over the IMF equals unity even if the other parameters change.

The reason for the seeming discrepancy between the theoretical expectation of a varying IMF and the lack of observational evidence for it is that the IMF of a single star-formation event is hardly ever observed directly. Instead, it has to be inferred from stellar populations that have already evolved. However, despite these difficulties, two distinct variations of the IMF at high stellar masses have been established in more recent times.

Firstly, the upper mass limit for stars, $m_{\rm max}$, correlates with the total mass of the stars that are created in a star-forming event, $M_{\rm ecl}$. This correlation is such that $m_{\rm max}$ is always much smaller than $M_{\rm ecl}$, even if $M_{\rm ecl}$ is below the physical mass limit for stars, $m_{\rm max *}$ \citep{Weidner2005,Weidner2006,Weidner2010,Yan2017}. Thus, for low-mass star clusters, $m_{\rm max} < m_{\rm max*} < M_{\rm ecl}$ and for high-mass star clusters $m_{\rm max} \approx m_{\rm max*} < M_{\rm ecl}$. This implies that massive stars do not form isolated.

Secondly, the high-mass slope of the IMF, $\alpha_3$, depends on the metallicity, $Z$, and the density, $\rho_{\rm ecl}$, of the star-forming molecular cloud \citep{Marks2012a}. The $\alpha_3$-$\rho_{\rm ecl}$ relation can also be transformed into a $\alpha_3$-$M_{\rm ecl}$ relation \citep{Marks2012b}.

Thus, both identified variations of the IMF depend on $M_{\rm ecl}$, i.e. the mass of the embedded star clusters.

The effort of quantifying the IGIMF as a function of the SFR has already been made by several authors ( e.g. \citealt{Weidner2011,Weidner2013b,Fontanot2017,Yan2017}). However, what is missing so far are simple interpolation formulae that can be used to quantify the IGIMF easily in early-type galaxies. This will be provided in this contribution, based on the tabulated values in \citet{Fontanot2017}, which parametrize the shape of the IGIMF as a function of the SFR. Moreover, the resulting IGIMF-parameters are expressed as functions of luminosity and stellar mass, instead of the SFR. For the latter step, the catalogue of early-type galaxies by \citet{Dabringhausen2016a} is used, and a relation established by \citet{Thomas2005} which links the internal velocity dispersions of the galaxies (listed in the catalogue) to the SFRs that are expected for them at the time when the majority of their stellar population formed, $SFR_{\rm peak}$.

This paper is organized as follows. Section~(\ref{sec:data}) describes the data on ETGs that is used for linking IGIMF-related parameters of ETGs to their observed properties. Section~(\ref{sec:methods}) introduces the IGIMF-model, including some adjustments and parametrisations of the IGIMF used in this paper. In Section~(\ref{sec:results}) high-mass IGIMF slopes and resulting masses of the ETGs are shown over their luminosities, and also interpolation functions of these parameters as functions of luminosity are presented. Numerous passbands are considered in this context. A discussion of the results is given in Section~(\ref{sec:discussion}), and a summary and conclusion in Section~(\ref{sec:conclusion}).

\section[Data]{Data}
\label{sec:data}

\subsection{Selection criteria for the sample of ETGs}
\label{sec:selection}

The data that is used to link the IGIMF to observed early-type galaxies is provided in the catalogue by \citet{Dabringhausen2016a}. This catalogue comprises 1715 ETGs, which span the whole luminosity range of ETGs from faint dwarf spheroidal galaxies (mostly from the catalogue by \citealt{McConnachie2012}) to giant elliptical galaxies (to a large extent from the ATLAS$^{\rm 3D}$ survey, \citealt{Cappellari2011}). What motivates to combine these galaxies into a single catalogue, and similar galaxies from other sources as well, is that they share two properties: There is little (if any) star formation in them at present, and random motion dominates over ordered motion for their stellar populations. Apart from these two defining properties, the properties of the ETGs are diverse. However, similar to main-sequence stars (e.g. \citealt{Demircan1991}), ETGs gather close to a (one-dimensional) line in the (two-dimensional) mass-radius space.

The quantities from the catalogue by \citet{Dabringhausen2016a} that are relevant for the present paper are the masses of their stellar populations under the assumption the the IMF is canonical, $M_{\rm can}$, and at least the Johnson-Cousins $V$-band luminosity, $L_V$. Also the luminosities in various other passbands are taken from that source if available, namely the $U$, $B$ and $I$-band in the Johnson-Cousins magnitude system and the $u$, $g$, $r$, $i$ and $z$-band in the SDSS magnitude system. To address several unspecified passbands of the above, the notation $L_{pb}$ is introduced, where the subscript $pb$ acts as a placeholder for "passband". Luminosities and stellar masses are key parameters for characterizing a galaxy, and at least luminosities are comparatively easy to measure. Luminosities and the stellar masses of the ETGs are thus selected as the variables in the interpolation functions of the IGIMF that are to be constructed here. The $V$-band luminosities are additionally needed for estimating $M_{\rm can}$ (see Section~\ref{sec:data-masses}).

As additional constraints, only galaxies for which also the central line-of-sight velocity dispersion $\sigma_0$, the spectroscopic age $t$ and the metallicity $Z$ are available from the catalogue by \citet{Dabringhausen2016a} were selected for the present paper. It is given for 460 ETGs and limits the data to the ETGs that were studied in greater detail. However, these {460} ETGs still cover the almost whole luminosity range of ETGs, from $L_V \approx 10^4 {\rm L}_{\odot}$ to $L_V \approx 10^{11} {\rm L}_{\odot}$.

As a caveat, note that the catalog by \citet{Dabringhausen2016a} merges data from many different teams, who obtained their raw data under different conditions with different instruments, and also reduced the raw data in different ways. As a consequence, the data in \citet{Dabringhausen2016a} is inherently inhomogeneous, even though some basic transformations to the data were applied where this seemed possible and prudent to alleviate this issue. The advantages of the catalogue by \citet{Dabringhausen2016a} is however that it contains all the data required in this work (i.e. $M_{\rm can}$, the $V$-band luminosity a spectroscopic age, and possibly other luminosities), and that it covers the whole luminosity range of ETGs, including faint objects that are not found in homogeneous samples from large galaxy surveys.

While the reader is referred to \citet{Dabringhausen2016a} for a detailed description of their catalogue, some basic information on the data in their catalogue that is relevant for the present paper is also given in the following.

\subsection{Luminosities}
\label{sec:data-luminosities}

The data on the luminosities (in different passbands) in the catalogue by \citet{Dabringhausen2016a} are either based on direct observations of the individual ETGs, or are derived from these data using their statistical properties. \citet{Dabringhausen2016a} prefer data based on direct observations, which they obtain from apparent magnitudes collected from many sources in the literature and the distance estimates they adopt for the according galaxies. They perform some basic homogenisation for this type of data, using overlaps between the different samples of galaxies introduced in the literature that they use. This is done in order to take care of offsets between the different data samples, which most likely originate from observations with different telescopes at different times, and different procedures for data reduction. 

If no published value for the apparent magnitude of a galaxy was available from the literature, while there was such data from neighbouring passbands, \citet{Dabringhausen2016a} use relations between luminosities of the same galaxies in different passbands to calculate the unknown luminosities from the measured luminosities. These relations are linear in $\log_{10}(L_{pb})$ and thus essentially in magnitudes. To estimate the uncertainties to the luminosities calculated with these relations, \citet{Dabringhausen2016a} compared the calculated values with the values from individual measurements where they were available. They derived the uncertainty from the scatter around the 1:1 relation with the observed luminosities on the $x$-axis and the calculated luminosities on the $y$-axis. Depending on the considered passband, they found values between $\log_{10}(L_{pb}/{\rm L}_{\odot})=0.17$ and $\log_{10} (L_{pb}/{\rm L}_{\odot})=0.20$. This is comparable to what is achieved with the better known formulae by \citet{Blanton2007}, where the conversion is done based on a colour and a luminosity, instead of up to two luminosities (see figure~(1) in \citealt{Dabringhausen2016a}). The advantage of the method by \citet{Dabringhausen2016a} is that it can be adopted easily to the number of other passbands that are available to calculate the luminosity of a given galaxy, while the method by \citet{Blanton2007} is restricted to exactly one color and one passband. For how many galaxies the luminosity is not directly observed, but calculated with an interpolation depends a lot on the passband: It is the majority of galaxies in the $I$-band, a bit less than one half in the $U$-band and the $V$-band, about 20 per cent in the $u$-band, the $g$-band and the $r$-band and less than 10 per cent in the $B$-band, the $i$-band and the $z$-band.

$L_V$ is listed in \citet{Dabringhausen2016a} for all ETGs for which they also list $R_{\rm e}$, $\sigma_{\rm 0}$ and $M_{\rm can}$. This means that in the $V$-band, the availability of the luminosity adds no further constraint on the ETGs selected from \citet{Dabringhausen2016a} for this paper. The situation is different with most other passbands, where data on luminosities other that in the $V$-band is missing especially for many low-mass ETGs (or dwarf Spheriodals). In consequence, the number of galaxies considered in this paper varies from passband to passband. Moreover, if the luminosity of a ETG is in a given passband inconsistent with its luminosity in the $V$-band, it is discarded from the sample considered in that passband, for reasons explained at the end of Section~\ref{sec:data-masses}. This further limits the number of ETGs considered in passbands other than the $V$-band, but happens comparatively rarely. One of the most severe cuts is in the $I$-band, where the data in the sample is decreased from 460 ETGs to 347 ETGs.
 
\subsection{Masses of the stellar populations} 
\label{sec:data-masses}

\begin{figure*}
\centering
\includegraphics[scale=0.85]{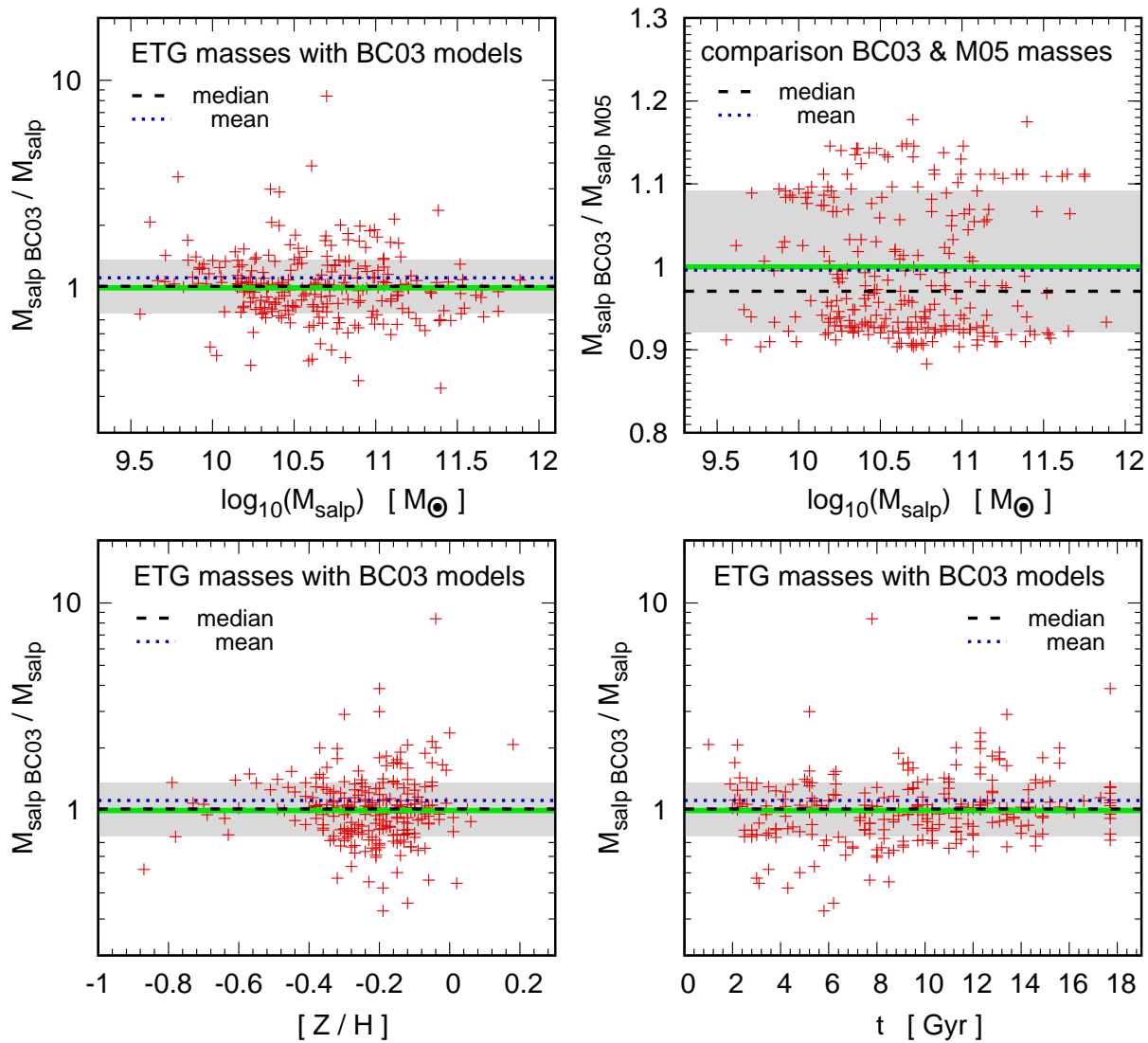}
\caption[A comparison between mass estimates based on a single SSP-model and mass estimates based on a sum of multiple SSP-models]{\label{fig:SSP-comp} A comparison between mass estimates based on a single SSP-model and mass estimates based on a sum of multiple SSP-models. The upper left panel shows the ratios between mass estimates from a single best-fitting SSP-model for each galaxy and mass estimates from linear combinations of multiple SSP-models, over the mass estimate from linear combinations of multiple SSP-models. The mass estimates from single SSP-models are obtained with the method described in \citet{Dabringhausen2016a} and based on the SSP-models by \citet{Bruzual2003}. The mass estimates from multiple SSPs are taken from \citet{Cappellari2013b}. The upper right panel shows for comparison the ratio between mass estimates for the ETGs with the method in \citet{Dabringhausen2016a} based on the SSP-models by \citet{Bruzual2003} and estimates with the same method based on the SSP-models by \citet{Maraston2005}. The lower left and the lower right panel are the same as the upper left panel, except that there is the metallicity instead of the stellar mass on the $x$-axis in the lower right panel, and the age in the lower left panel.}
\end{figure*}

The estimates on $M_{\rm can}$ given in \citet{Dabringhausen2016a} are based on a large set of models for simple stellar populations (SSPs) by \citet{Bruzual2003}. These SSPs are defined as stellar populations that have formed instantly at a certain time with a certain metallicity. More specifically, \citet{Dabringhausen2016a} obtain $M_{\rm can}$ of a given ETG by first searching an SSP-model by \citet{Bruzual2003} that represents the age and the colours of the ETG the best, then adopting the $M/L_V$ predicted by the according SSP-model as the $M/L_V$ of the ETG, and finally multiplying this $M/L_V$ by the $L_V$ of the ETG. The colours of the ETGs serve in this context as indicators for the metallicities of their stellar populations, since the metallicity determines the colours of an SSP with a given age, while colours are much easier to observe than metallicities.

The data on the ages and the colours of the stellar populations of the ETGs that \citet{Dabringhausen2016a} use in the estimates of $M_{\rm can}$ come from the literature that they consider.

There are certainly more elaborate methods to determine the stellar mass $M$ of a galaxy than the method in \citet{Dabringhausen2016a}. Instead of searching for a single SSP-model that represents the age and the colour of a given ETG the best, they are based on reconstructing the observed spectrum of the galaxy as a sum of SSP-models, where the coefficient of every SSP-model contributing to the spectrum has to be $\ge 0$ (e.g. \citealt{Blanton2007,Cappellari2013b}). A motivation to consider this method for ETGs is that they do not form instantly. There is also a tendency that the star formation takes longer in less massive ETGs \citep{Thomas2005,delaRosa2011}. In some low-mass ETGs in the Milky Way (also known as Milky Way dwarf Spheroidals), star formation seems to be a process that continues almost over the whole age of the Universe \citep{Weisz2014}. Traces of fairly recent star formation and the presence of dust in ETGs were also reported by \citet{Schawinski2007a} and \citet{Schawinski2007b}.

Simpler methods for estmating the $M$ of ETGs (dwarfs as well as giants) are however not uncommon; also in the rather recent literature. For instance, \citet{Forbes2008}, \citet{Forbes2011} and \citet{Misgeld2011} consider the mass-to-light ratios of ETGs as functions of colour from SSPs with a single age. The colours thereby become proxies for metallicity while a possible presence of dust is not considered in their approach. \citet{Taylor2011} argue that already visual colors are by themselves (i.e. without near-infrared colors) are a pretty good indicator for stellar mass-to-light ratios. Finally, \citet{McConnachie2012}, whose catalogue is the source for the least luminous ETGs in \citet{Dabringhausen2016a}, assumes a mass-to-light ratio of one in Solar units for all galaxies in his sample. This is equivalent to assuming that all galaxies in his sample have formed with only one SSP without variations in age and metallicity.

The approach taken in \citet{Dabringhausen2016a}, and thus here as well, is inspired by \citet{Misgeld2011}, see their equations~(1) to~(6). However, unlike \citet{Misgeld2011}, \citet{Dabringhausen2016a} did not assume the same age for all ETGs. Instead, they defined a grid of fitting functions for different ages, and approximated the stellar mass of each galaxy with the function that was closest to its spectroscopic age, if a spectroscopic age was available. In the present paper, only those galaxies with a spectroscopic age are considered. With the age at least roughly given, metalicity is the remaining coordinate that determines the metallicity. Thus, the age-metalicity degeneracy is broken to some extent, and the approach taken here is already an improvement over the afore mentioned efforts.

Regarding dust and recent star formation in ETGs, note that \citet{Schawinski2007a} state that their sample contains some ETGs with a $(NUV-r)$ colour above 6.5, which they say cannot be explained without the presence of dust, and a sizable fraction of ETGs with $(NUV-r)$ colours below 5.4, which they take as an indicator for recent star formation. According to \citet{Schawinski2007b}, recent star formation took place in 20 per cent of their sample of ETGs, which formed 1 to 10 percent of their stellar mass some 100 Myr ago. This means on the other hand that most ETGs in their sample do not show signs of recent star formation, and even in those which do, most stars are old. Thus, most ETGs in the sample used in \citet{Schawinski2007a} and \citet{Schawinski2007b} end up in between the two extremes in $(NUV-r)$ colour, which means that they may or may not have stars recently and contain dust or not. However, they can also be old and gas-free from their colours.

The key question in the context of the present paper is whether the comparatively simple estimates of the stellar mass of the ETGs in \citet{Dabringhausen2016a} are sufficent for the purpose here, or whether more elaborate estimates like in \citet{Blanton2007} or \citet{Cappellari2013b} are required. The improvement that may be expected is illustrated in Fig.~(\ref{fig:SSP-comp}), where the stellar masses derived from a single SSPs are compared to estimates for the stellar masses of the same galaxies with multiple SSPs. The mass estimates based on multiple stellar populations come from \citet{Cappellari2013b} and comprise the ATLAS$^{3D}$ sample of ETGs. The mass estimates from the single SSP model are calculated from the mass-to-light ratio in the $V$-band with
\begin{equation}
\label{eq:MLcolour}
\frac{M}{L_{V}}=a \, \arctan ( b \, \frac{1}{2}[(g-r)+(g-i)] + c ) + d,
\end{equation}
where $(g-r)$ and $(g-i)$ are the colours of the galaxy in the SDSS-system and the parameters $a$, $b$, $c$, and $d$ are evaluated for 3, 5, 7, 9, 11 and 13 Gyr. For each galaxy, the age that lies closest to its actual age is chosen to compute its mass-to-light ratio. Then it is multiplied by its luminosity to obtain its stellar mass. Note that the age-metallicity degeneracy is formally dissolved for single ages of the ETGs, since ages are given for all ETGs and the colours can then be translated into a metallicity. Also note that the stellar masses are calculated like in \citet{Dabringhausen2016a}, but with the coefficients $a$, $b$, $c$ and $d$ in their equation~(18) adjusted to the Salpeter IMF instead of the canonical IMF. The latter modification is necessary since also the mass estimates in \citet{Cappellari2013b} are based on the assumption of the Salpeter IMF.

The upper left panel and the two lower panels of Fig.~(\ref{fig:SSP-comp}) show that the difference between the mass estimate from a single SSP and the mass estimate from a sum of multiple SSPs varies strongly from galaxy to galaxy, and is very substantial (a factor of a few) for some galaxies. The upper right panel of Fig.~(\ref{fig:SSP-comp}) shows the ratios between the mass estimates with the SSP-models by \citet{Bruzual2003}, which is the default in this paper, and \citet{Maraston2005}. All estimates in that panel are made with equation~18 in \citet{Dabringhausen2016a}, and thus based on finding a single best-fittng SSP for each ETG. The difference between the two estimates is in that case never larger than 20 per cent. Thus, the spread in the mass ratios caused by using multiple instead of single SSPs for the fits is indeed much larger than the spread caused by going from one set of SSP models to another.

On the other hand, the masses estimated with a single SSP deviate from the masses estimated with multiple SSPs by about 30 per cent at most for about two thirds of the galaxies, and actually considerably less in many cases. The mean of the ratio between the estimates with a single SSP and the estimates with multiple SSPs is approximately 1.1; i.e. the estimates from multiple SSPs are on average 10 per cent lower than the estimates with a single SSP. This is to a large extent driven by some extreme outliers, and thus the median of the ratios is almost identical to 1. Relevant for the purpose in this paper are such average relations, and not the precise masses of individual galaxies. The average decrease of the mass estimates for the ETGs by changing from single SSP-models to multiple SSP-models in fact for many galaxies similar to the change from one set of SSP-models to another in many ETGs (i.e approximately 10 per cent, see upper right panel of Fig.~\ref{fig:SSP-comp}). The effect that the IGIMF is expected to have on the mass estimates of galaxies is more pronounced, and may amount to a factor of two for the average masses of the most luminous galaxies (see Section~\ref{sec:results}). It thus appears acceptable to consider mass estimates from single SSP-models in the present context, given that the intent of this paper is to keep matters simple, at the expense that some issues are not covered to the greatest possible detail. The main idea is to give readers at least a rough idea by which factor an estimate of the stellar mass based on the canonical IMF should be corrected in order to capture the effect of the IGIMF on the actual stellar mass.

As a caveat, note that Figure~(\ref{fig:SSP-comp}) covers with the ATLAS$^{3D}$ only ETGs in mass range from $\log_{10}(M/{\rm M}_{\odot})=9.5$ to  $\log_{10}(M/{\rm M}_{\odot})=12$ for the Salpeter IMF, and a bit lower values for the canonical IMF. This means that estimates for $M$ of dwarf galaxies are not tested in Fig.~(\ref{fig:SSP-comp}). Note however that Fig.~(\ref{fig:SSP-comp}) does not suggest a notable trend of the ratios between $M$ estimated with a single SSP or $M$ estimated with multiple SSPs for each galaxy with mass (upper left panel), metallicity (lower left panel) or age (lower right panel).

Also note that in estimates of $M_{\rm can}$, the stellar initial mass function is set to the canonical IMF with $m_{\rm max} = 100 {\rm M}_{\odot}$ (cf. eq~\ref{eq:IMF}). However, as already noted in Section~\ref{sec:introduction}, the actual distribution of stellar stellar masses in galaxies are not given by the canonical IMF, but by the IGIMF. In order to distinguish the real stellar masses according to the IGIMF model from $M_{\rm can}$, the former will be denoted as $M_{\rm IGIMF}$ in the present paper.

A last caveat is that the estimates for $M_{\rm can}$ are only based on the values for $L_V$ of the ETGs. If the ETGs were SSPs, the estimates for their $M_{\rm can}$ should in principle be the same no matter which passband is used to estimate them from the $M/L$-ratios predicted by SSP-models. However, in reality ETGs are not SSPs (even if their stellar populations formed rapidly), and the measurements of the luminosities in different passbands are subject to observational uncertainties. For some galaxies, the data on the luminosities in different passbands even cannot be explained with any realistic stellar population. For practical reasons, $L_V$ is in such cases assumed to be the standard and if the luminosity in some other passband $pb$, $L_{pb}$, is inconsistent with it, the ETG is discarded from the sample in this passband. The criterion for rejection is that the value for $M_{\rm can}/L_{pb}$ calculated for the ETG in question is more than 20 percent above the $M/L_{pb}$-ratio for a 15 Gyr old SSP with a metallicity of $[\rm{Z}/\rm{H}]=0.66$ according to the SSP-models by \citet{Maraston2005}. These SSP-models have been chosen because they provide with their extreme ages and metallicities generous but not arbitrary upper limits for phyically plausible $M/L_{pb}$. The underlying stellar models for these SSP-models have been calculated by \citet{Salasnich2000}.

\section[Methods]{Methods}
\label{sec:methods}

\subsection{The shape of the IGIMF in dependence of the SFR}
\label{sec:IGIMF-SFR}

The fundamental underlying assumption for the IGIMF is that most, if not all stars form in groups and not in isolation. This notion is well supported by observations (e.g. \citealt{Lada2003}), and the resulting groups of stars are called 'embedded clusters' as long as they are still located inside the gas cloud out of which they formed. Most of these groups may disperse shortly after their formation and would thus not become long-lived star clusters, either because the stars become unbound through the expulsion of the residual gas \citep{Lada2003,Fall2005,Goodwin2006}, or because tidal fields may destroy the embedded cluster \citep{Kruijssen2012}. However, important in the context of the IGIMF of a galaxy is only that each embedded cluster is characterized by a specific IMF, while the actual fate of the embedded clusters is irrelevant.

There are two variations of the shape of the IMF that have to be considered. Both of them depend on the total mass of the stellar population of the embedded cluster, $M_{\rm ecl}$.

The first type of variation of the IMF is a non-trivial dependence of the mass of the most massive star that formed in an embedded cluster on $M_{\rm ecl}$, as illustrated in figure~(2) in \citet{Weidner2010} and figure~(1) in \citet{Yan2017}. This dependence has been formulated in \citet{Pflamm2007} through equations
\begin{equation}
\label{eq:Mecl}
M_{\rm ecl} = k_{\rm ecl} \int^{m_{\rm max}}_{m_{\rm low}} m \, \xi_{\rm IMF} (m) \, dm
\end{equation} 
and
\begin{equation}
\label{eq:Mmax}
1 = k_{\rm ecl} \int^{m_{\rm max*}}_{m_{\rm max}} \, \xi_{\rm IMF} (m) \, dm,
\end{equation} 
where $M_{\rm ecl}$ is the total mass of the stars in the embedded cluster, $\xi_{\rm IMF} (m)$ is the initial stellar mass function, $k_{\rm ecl}$ is a normalisation that ensures that the integration on the right side of equation~(\ref{eq:Mecl}) indeed returns $M_{\rm ecl}$, $m_{\rm low} \approx 0.1 \ {\rm M}_{\odot}$ is the mimimum mass for stars, $m_{\rm max}$ is the mass of the most massive star in the cluster and $m_{\rm max*}$ is the physical mass limit for stars, which according to \citep{Weidner2004a} is around 150 ${\rm M}_{\odot}$. A parametrisation of the dependence of $m_{\rm max}$ on $M_{\rm ecl}$ is given in \citet{Pflamm2007} as
\begin{align}
\log_{10}(m_{\rm max})  & = 2.56 \log_{10}(M_{\rm ecl}) \nonumber \\ 
& \times (3.82^{9.17} + [\log_{10} (M_{\rm ecl})]^{9.17})^{(1/9.17)}-0.38,
\label{eq:Mmaxfit}
\end{align}
which has been obtained through a fit to numerical solutions to equations~(\ref{eq:Mecl}) and~(\ref{eq:Mmax}). According to equation~(\ref{eq:Mmaxfit}), the expected $m_{\rm max}$ is much lower than $M_{\rm ecl}$ even in very small embedded clusters with $M_{\rm ecl}<m_{\rm max*}$. Figure~(2) in \citet{Weidner2010} illustrates that the IMF-variation formulated with equations~(\ref{eq:Mecl}) to~(\ref{eq:Mmaxfit}) reflects the observations quite well.

The second type of variation of the IMF is a variation of the high-mass slope, $\alpha_3$, of the IMF (cf. equation~\ref{eq:IMF}), especially in very massive embedded clusters. While \citet{Marks2012a} find that $\alpha_3$ is also a function of the metallicity of the embedded cluster, the dominant dependency is clearly the one on the stellar density, $\rho_{\rm ecl}$. It is given in \citet{Marks2012a} as
\begin{equation}
\label{eq:alpha3}
\alpha_3=
\begin{cases}
2.3 & {\ \rm if \ } \rho_{\rm ecl} < 95000 \, {\rm M}_{\odot}/{\rm pc^3}\\ 
7.86-0.43 \, \log_{10}(\rho_{\rm ecl}) & {\ \rm if \ } \rho_{\rm ecl} \ge 95000 \, {\rm M}_{\odot}/{\rm pc^3}.
\end{cases}
\end{equation}
\citet{Marks2012b} show by an analysis of binary populations that $M_{\rm ecl}$ and $\rho_{\rm ecl}$ are correlated such that
\begin{equation}
\label{eq:rhoMecl}
\log_{10} (\rho_{\rm ecl}) = 0.61 \, \log_{10}(M_{\rm ecl})+2.85,
\end{equation}
where $M_{\rm ecl}$ is measured in ${\rm M}_{\odot}$ and $\rho_{\rm ecl}$ in ${\rm M}_{\odot} \, pc^{-3}$. Thus, just like the variation of $m_{\rm max}$, the variation of $\alpha_3$ can be expressed as a function of $M_{\rm ecl}$.

Crucial for the IGIMF of a galaxy is then the mass spectrum of embedded clusters that form in it, which is quantified with the embedded cluster mass function (ECMF). \citet{Lada2003} find that the ECMF can be formulated as
\begin{equation}
\label{eq:ECMF}
\Xi(M_{\rm ecl}) \propto M_{\rm ecl}^{\beta}
\end{equation}
with $\beta=2$. A similar set of equations like equations~(\ref{eq:Mecl}) and~(\ref{eq:Mmax}) for star clusters also sets the upper mass limit for star clusters that form in a galaxy, namely
\begin{equation}
\label{eq:MSFR}
\delta t \times {\rm SFR} = k_{\rm SFR} \int^{M_{\rm ecl,max}}_{M_{\rm ecl,low}} m \, \Xi (M_{\rm ecl}) \, dM_{\rm ecl}
\end{equation} 
and
\begin{equation}
\label{eq:Meclmax}
1 = k_{\rm SFR} \int^{M_{\rm ecl,up}}_{m_{\rm ecl,max}} \, \Xi (M_{\rm ecl}) \, dM_{\rm ecl},
\end{equation} 
where $\delta t$ is a characteristic time in which a representative population of star clusters forms in a given galaxy, SFR is the star formation rate, $k_{\rm SFR}$ is a normalisation constant that ensures that the integration on the right side of equation~(\ref{eq:MSFR}) returns the total mass of all star clusters that form during $\delta t$, $M_{\rm ecl,max}$ is the mass of the most massive star cluster in the considered galaxy, $M_{\rm ecl,low}$ is the lower mass limit for star clusters and $M_{\rm ecl,up}$ is the absolute upper mass limit for star clusters \citep{Kroupa2013}. The value of $\delta t$ is about 10 Myr, independent of the SFR \citep{Weidner2004b}, and \citet{Fontanot2017} use $M_{\rm ecl,low} = 5 \, {\rm M}_{\odot}$ and $M_{\rm ecl,up} = 2 \times 10^7 \, {\rm M}_{\odot}$ in their calculations. While equations~(\ref{eq:MSFR}) and~(\ref{eq:Meclmax}) could be used to calculate the dependency of $M_{\rm ecl,max}$ on the SFR, \citet{Weidner2004b} have already found
\begin{equation}
\label{eq:MaxClusterMass}
\log_{10}(M_{ecl,max})=0.746 \, \log_{10} (SFR) + 4.93
\end{equation}
from observational data. Thus, the SFR is taken as the key parameter that determines the IGIMF of a galaxy, which is given as
\begin{align}
\xi_{\rm IGIMF}(m) & = \int^{M_{\rm ecl,max}{\rm (SFR)}}_{M_{\rm ecl,min}} \xi_{\rm IMF} (m \le m_{\rm max}(M_{\rm ecl})) \nonumber \\
& \times \Xi_{\rm ecl} (M_{\rm ecl}) \, dM_{\rm ecl},
\label{eq:IGIMF1}
\end{align}
i.e. essentially the summation of the stellar populations in all embedded clusters having formed in a time interval $\Delta t$ \citep{Weidner2005}. The reasons for a correlation between the SFR and the shape of the IGIMF are poorly understood, but it would make sense that the SFR alters the properties of the star-forming interstellar medium, like turbulence, temperature or metal enrichment. This may influence the mass spectrum of newly formed stars (see \citealt{Weidner2013a}).

\citet{Fontanot2017} use equation~(\ref{eq:IGIMF1}) with equations~(\ref{eq:IMF}),~(\ref{eq:Mmaxfit}),~(\ref{eq:alpha3}),~(\ref{eq:rhoMecl}),~(\ref{eq:ECMF}) and~(\ref{eq:MaxClusterMass}) to calculate the IGIMF in galaxies in dependence of their SFR. They find that the shape of the resulting IGIMF can well be approximated as
\begin{equation}
\xi_{\rm IGIMF} (m) =k \, k_i \, m^{-\alpha_i},
\label{eq:IGIMF2}
\end{equation}
with 
\begin{align}
\nonumber \alpha_1 = 1.3  & {\ \rm if \ } 0.1  \le  \frac{m}{\rm{M}_{\odot}} < 0.5,\\
\nonumber \alpha_2 = 2.3  & {\ \rm if \ } 0.5  \le  \frac{m}{\rm{M}_{\odot}} < m_1,\\
\nonumber \alpha_3 \, \in \, \mathbb{R}  & {\ \rm if \ } m_1 \le  \frac{m}{\rm{M}_{\odot}} < m_{\rm break}, \\
\nonumber \alpha_4 \, \in \, \mathbb{R}  & {\ \rm if \ } m_{\rm break} \le  \frac{m}{\rm{M}_{\odot}} \le m_{\rm max},
\end{align}
for all considered SFRs. In this equation, $m$ is the initial stellar mass, $m_1$ is a mass close to $1 \, {\rm M}_{\odot}$, $m_{\rm break}$ is a mass in the range of intermediate to high-mass stars, $m_{\rm max}$ is the maximum initial stellar mass, the factors $k_i$ ensure that the IGIMF is continuous where the power changes and $k$ is a normalization constant that ensures that the integral over the IGIMF equals unity, even if the other parameters change.  Note that equation~(\ref{eq:IGIMF2}) is conceptually very different from equation~(\ref{eq:IMF}) despite its structural similarity. Equation~(\ref{eq:IMF}) parametrizes the mass spectrum of newly formed stars in individual star forming regions within a galaxy, while equation~(\ref{eq:IGIMF2}) parametrizes the overall mass spectrum of all stars born in the many star forming regions of the galaxy. The parameters $m_1$, $\alpha_3$, $m_{\rm break}$, $\alpha_4$ and $m_{\rm max}$ are dependent of the SFR, and are tabulated for selected values of the SFR in table~(1) in \citet{Fontanot2017}.

The parametrizations of the IGIMF and its impact on the properties of early-type galaxies provided in this paper are based on the values listed in table~(1) in \citet{Fontanot2017}, which means that they implicitly rely on the same assumptions and relations that \citet{Fontanot2017} use.

As a caveat, note that Equations~(\ref{eq:IGIMF1}) and~(\ref{eq:IGIMF2}) treat the IGIMF as a galaxy-wide property of the stellar population. However, in real galaxies, the SFR is probably a more local parameter. It most likely depends on the density of the interstellar medium (ISM). \citet{MartinNavarro2015} and \citet{vanDokkum2017} have indeed discovered gradients in the stellar mass functions of massive ETGs, so that the departure from the canonical IMF is the strongest in the central parts of these ETGs. The interpretation of this finding in the IGIMF model would be that the SFR was the highest in these central parts the the galaxies, where also the density is the highest. This also implies that, even there were an exact dependency between the SFR and the shape of the IGIMF, two galaxies with the same galaxy-wide SFR could have different IGIMFs if their local SFRs were different.

\subsection{Simplifing the parametrization of the IGIMF}
\label{sec:simplification}

\begin{figure}
\centering
\includegraphics[scale=0.85]{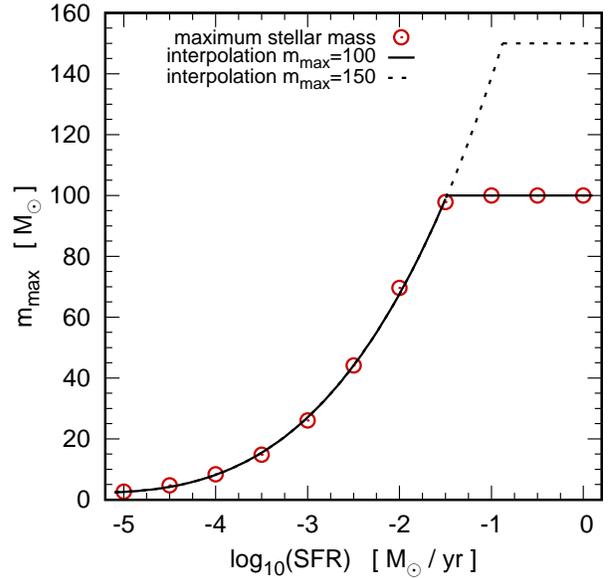}
\caption[The most massive stars formed in a galaxy in dependence of the SFR.]{\label{fig:SFR-mmax} The mass of the most massive stars that form in a galaxy in dependency of its SFR. The (red) circles show calculations by \citet{Fontanot2017}, and the (black) solid line is a fit to them, which is given by equation~\ref{eq:interpolationMmax}. The  black dashed line shows an adaption of equation~(\ref{eq:interpolationMmax}) to an upper physical mass limit of $m_{\rm max*}=150 \ {\rm M}_{\odot}$ instead of $m_{\rm max*}=100 \ {\rm M}_{\odot}$.}
\end{figure}

\begin{figure*}
\centering
\includegraphics[scale=0.85]{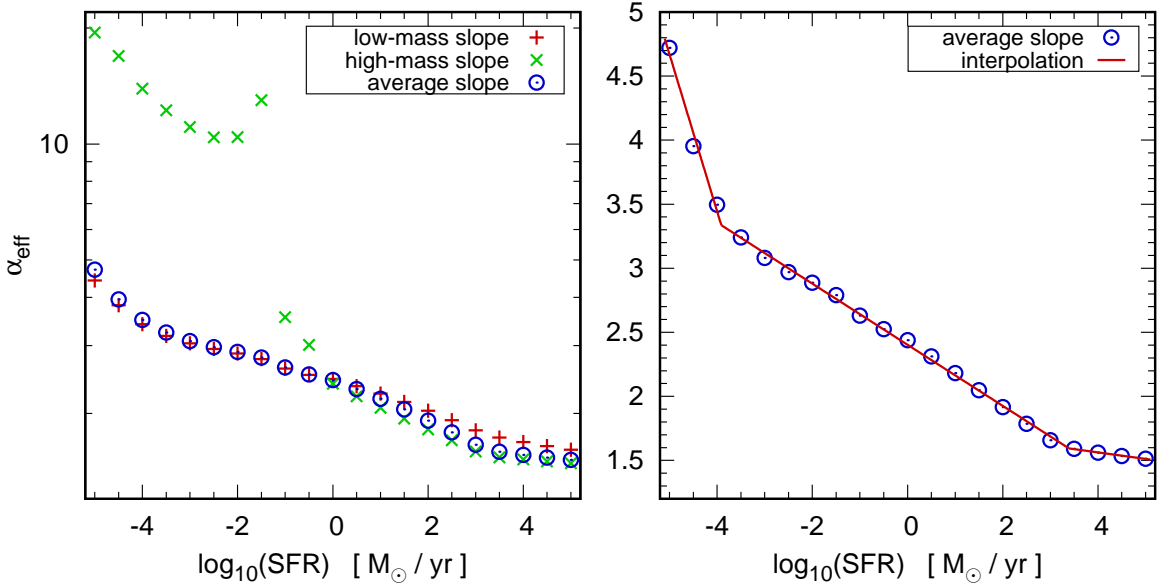}
\caption[The effective high-mass slope of the IGIMF as a function of the SFR.]{\label{fig:SFR-alpha_eff} The effective high-mass slope of the IGIMF, $\alpha_{\rm eff}$, as a function of the SFR. The left panel shows $\alpha_{\rm eff}$ as blue circles, as well as the values for $\alpha_3$ (red plusses) and $\alpha_4$ (green x-symbols) in equation~(\ref{eq:IGIMF2}), from which $\alpha_{\rm eff}$ is calculated with equation~(\ref{eq:alphaeff}). The right panel shows the values for $\alpha_{\rm eff}$ and an interpolation to them, which is given by equations~(\ref{eq:alphaeff21}) to~(\ref{eq:alphaeff23}) in their respective regimes.}
\end{figure*}

According to equations~(\ref{eq:Mmaxfit}) and~(\ref{eq:MaxClusterMass}), the maximum mass of stars that form in a galaxy depend on the SFR. In consequence, the IGIMF is truncated below the physical mass limit for stars if the SFR is very low.

The according data in table~1 in \citet{Fontanot2017} for low SFRs can be parametrized well as
\begin{align}
\label{eq:interpolationMmax}
\nonumber \frac{m_{\rm max}}{{\rm M}_{\odot}} = & \left[\log_{10}\left(\frac{SFR}{{\rm M}_{\odot}{\rm yr}^{-1}}\right)+5.82\right]^{3.13}\\
- & 0.409\times \log_{10} \left(\frac{SFR}{{\rm M}_{\odot}{\rm yr}^{-1}}\right)
\end{align}
for $\log_{10} (SFR/{\rm M}_{\odot}{\rm yr}^{-1}) <-1.5$ if $m_{\rm max*}=100 \ {\rm M}_{\odot}$, or $\log_{10} (SFR/{\rm M}_{\odot}{\rm yr}^{-1})<-0.878$ if $m_{\rm max*}=150 \ {\rm M}_{\odot}$. This is illustrated in figure~(\ref{fig:SFR-mmax}). If the $SFR$ is higher than the respective threshold value, the most massive stars can form in the respective galaxy, i.e. $m_{\rm max}=m_{\rm max*}$.

The value of 100 ${\rm M}_{\odot}$ adopted by \citet{Fontanot2017} for $m_{\rm max*}$ is rather low, but consistent with the settings in the widely used simple stellar population (SSP) models by \citet{Bruzual2003} and \citet{Maraston2005}. Observational data suggest $m_{\rm max*} \sim 150 \ {\rm M}_{\odot}$ (e.g. \citealt{Weidner2004a,Oey2005}) or even higher \citep{Crowther2010}. However, in practice, the overall properties of a stellar population are rather insensitive to the exact value of $m_{\rm max*}$ due to the power-law nature of the IMF. An exception are very massive star clusters with extremely top-heavy IMFs, which may form however at very high SFRs.

In order to simplify the formulation of the IGIMF proposed by \citet{Fontanot2017}, the slopes $\alpha_3$ and $\alpha_4$ in equation~(\ref{eq:IGIMF2}) are combined to a single effective slope $\alpha_{\rm eff}$. This is done by calculating the arithmetic mean of $\alpha_3$ and $\alpha_4$, weighted by the mass that stars in the according mass range contribute to the overall stellar population. Thus, $\alpha_{\rm eff}$ is given as
\begin{equation}
\label{eq:alphaeff}
\alpha_{\rm eff}=\frac{M_{\alpha 3}\,\alpha_3+M_{\alpha 4}\, \alpha_4}{M_{\alpha 3}+M_{\alpha 4}}
\end{equation}
with
\begin{equation}
M_{\alpha_3} = \int^{m_{\rm break}}_{m_1} m\, \xi_{\rm IGIMF}(m) \, dm 
\end{equation}
and
\begin{equation}
M_{\alpha_4} = \int^{m_{\rm max}}_{m_{\rm break}} m\, \xi_{\rm IGIMF}(m) \, dm. 
\end{equation}
The parameter $m_1$ introduced by \citet{Fontanot2017} is also slightly dependent on the SFR, but always close to ${1\, \rm M}_{\odot}$. Therefore, $m_1=1\, {\rm M}_{\odot}$ is assumed here for simplicity.

The left panel of figure~(\ref{fig:SFR-alpha_eff}) shows $\alpha_{\rm eff}$ in comparison to $\alpha_3$ and $\alpha_4$ in equation~(\ref{eq:IGIMF2}). It illustrates that $\alpha_{\rm eff}$ is quite similar to $\alpha_{\rm 3}$, either because the slope at the highest stellar masses, i.e. $\alpha_4$, is very steep (at low SFRs) or very similar to $\alpha_3$ (at high SFRs).

The effective high-mass IGIMF slope can well be parametrised as a three-part linear function of $\log_{10} (SFR)$, namely
\begin{equation}
\label{eq:alphaeff21}
\alpha_{\rm eff}=-1.225\log_{10}\left(\frac{SFR}{{\rm M}_{\odot}{\rm yr}^{-1}}\right) -1.456
\end{equation}
 if $\log_{10}(SFR/{\rm M}_{\odot}{\rm yr}^{-1}) \le -3.91$,
\begin{equation}
\label{eq:alphaeff22}
\alpha_{\rm eff}=-0.239 \log_{10}\left(\frac{SFR}{{\rm M}_{\odot}{\rm yr}^{-1}}\right) + 2.402
\end{equation}
if $-3.91 < \log_{10}(SFR/{\rm M}_{\odot}{\rm yr}^{-1}) \le 3.39$, and
\begin{equation}
\label{eq:alphaeff23}
\alpha_{\rm eff}=-0.051 \log_{10}\left(\frac{SFR}{{\rm M}_{\odot}{\rm yr}^{-1}}\right) + 1.765
\end{equation}
if $\log_{10}(SFR/{\rm M}_{\odot}{\rm yr}^{-1}) > 3.39$, as the right panel of figure~(\ref{fig:SFR-alpha_eff}) shows. Arguably the most relevant of the above three equations, at least for the purpose here, is equation~(\ref{eq:alphaeff22}). Equation~(\ref{eq:alphaeff21}) is only relevant for low-mass ETGs with very steep high-mass IGIMFs, and it can be argued that it hardly matters whether equation~(\ref{eq:alphaeff21}) predicts $\alpha_{\rm eff} \approx 6$ for a galaxy or equation~(\ref{eq:alphaeff22}) predicts $\alpha_{\rm eff} \approx 4$ for the same galaxy, since massive stars are almost non-existent in it in any case. Also note that equation~(\ref{eq:alphaeff21}) is only relevant for galaxies where the formation time scales are so long that they may be problematic for the single-starburst approximation (see Section~\ref{sec:linking}). Equation~(\ref{eq:alphaeff22}) on the other hand is only relevant for the most extreme star formation events, which barely concerns ETGs to the very high-mass end. However, there may be situations outside the modelling of ETGs, where equations~(\ref{eq:alphaeff21}) and~(\ref{eq:alphaeff23}) are more useful.

Thus, in summary the simplified parametrisation of the IGIMF used in this the reminder of this paper is given as
\begin{equation}
\xi_{\rm IGIMF} (m) =k \, k_i \, m^{-\alpha_i},
\label{eq:IGIMF3}
\end{equation}
with 
\begin{align}
\nonumber \alpha_1 = 1.3  & {\rm \ if \ } 0.1 \le  \frac{m}{\rm{M}_{\odot}} < 0.5,\\
\nonumber \alpha_2 = 2.3  & {\rm \ if \ } 0.5 \le  \frac{m}{\rm{M}_{\odot}} < 1.0,\\
\nonumber \alpha_{\rm eff} \, \in \, \mathbb{R}  & {\ \rm if \ } 1.0 \le  \frac{m}{\rm{M}_{\odot}} \le m_{\rm max},
\end{align}
where $m$ is the initial stellar mass, $m_{\rm max}$ is a function of the SFR that is given by equation~(\ref{eq:interpolationMmax}) for $\log_{10}(SFR/{\rm M}_{\odot}{\rm yr}^{-1})<-0.878$ and by $m_{\rm max}=150 \ {\rm M}_{\odot}$ for $\log_{10}(SFR/{\rm M}_{\odot}{\rm yr}^{-1}) \ge -0.878$, $\alpha_{\rm eff}$ is a function of the SFR given by equations~(\ref{eq:alphaeff21})~to~(\ref{eq:alphaeff23}) depending on the SFR-range, the factors $k_i$ ensure that the IGIMF is continuous where the power changes and $k$ is a normalization constant that ensures that the integral over the IGIMF equals unity, even if the other parameters change. Despite the conceptual differences between IMF and IGIMF, we adopt the convention for the IMF by referring to an IGIMF as top-heavy if $\alpha_{\rm eff} < 2.3$ and as top-light if $\alpha_{\rm eff} > 2.3$.

Note that massive ETGs have in this picture top-heavy IGIMFs, whereas some authors (e.g. \citealt{LaBarbera2013}) find spectroscopic evidence for bottom heavy IGIMFs in the central regions of massive ETGs. This is only a problem at first sight, since top heaviness and bottom heaviness occur at different parts of the mass spectrum and possibly have different origins. Thus, in principle, a massive ETG can be both (see \citealt{Jerabkova2018} and the end of Section~\ref{sec:discussion}). However, bottom heaviness is according to \citet{Marks2012a} and \citet{Jerabkova2018} primarily a metallicity effect, which is not considered here.

\subsection{Linking the IGIMF to observed parameters of early-type galaxies}
\label{sec:linking}

\begin{figure}
\centering
\includegraphics[scale=0.85]{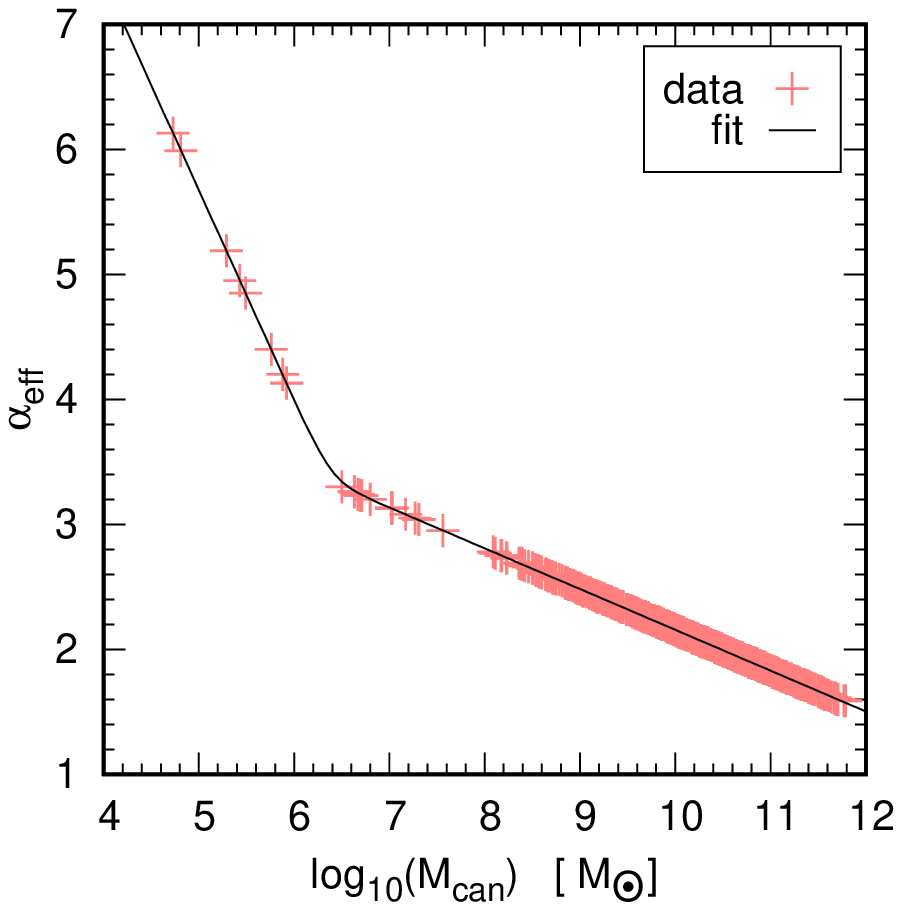}
\caption[\alpha_{\rm eff} in dependency of the mass of the galaxy.]{\label{fig:Mdyn-alpha} The effective high-mass slope of an ETG during its active time of star formation in dependency of its stellar mass assuming the canonical IMF, $M_{\rm can}$. $\alpha_{\rm eff}$ depends on the SFR, which is estimated from equation~(\ref{eq:deltatMdyn}), and $M_{\rm can}$ is taken from \citet{Dabringhausen2016a}. The (red) crosses show the resulting $\alpha_{\rm eff}$ for the case that the SFR is constant for the whole period $\Delta t$ and the black lines is the interpolation to the data.}
\end{figure}

While the SFR correlates with the shape of the IGIMF in galaxies, it is not the most practical parameter for estimating a characteristic IGIMF in early-type galaxies (ETGs) from directly observable quantities. The reason is that the present-day SFR in ETGs is so low that the present-day masses of their stellar populations imply that the SFR must have been much higher when the majority of their stars have formed. However, the abundance of $\alpha$-elements can serve as an indirect indicator for the SFR of an ETG in the past. The underlying notion is that different types of supernovae reinserted on different timescales different mixtures of elements into the interstellar medium (ISM), from which new stars were still forming. Type-II supernovae, which have high yields of $\alpha$-elements, are thought to be the final stage of the evolution of high-mass stars and therefore occur on a timescale of Myr after the formation of their progenitor stars. Type-Ia supernovae on the other hand, which have high yields of iron, are thought to be white dwarfs that surpass the Chandrasekhar-mass by accreting matter. Depending on the parameters of the progenitor binary system, it can take individual binaries Gyrs until one of the components becomes a SNIa, while the peak in the SNIa rate in a typical ETG is at a timescale of approximately 0.3 Gyr according to \citet{Matteucci2001}. The key point is however that the delay for SNIas in a galaxy is in any case longer than the delay for SNIIs. This is because SNIIs are linked to the timescale for the evolution of massive stars while SNIas require the remnant of an intermediate or low mass star. As a consequence, the SNIIs associated to a given star formation event naturally precede the SNIas associated to the same event. Long-lasting star formation leads however to a continung production of both SNII and SNIa progenitors, and shifts the peaks for both rates to higher ages of the ETGs. This suggest that the higher the $\alpha$-abundance in a stellar population of a ETG is in comparison to its iron-abundance, the sooner the ETG must have stopped to form stars.

On this basis, \citet{Thomas2005} have estimated the timescales for the formation of ETGs of different mass. They find
\begin{equation}
\log_{10} (\Delta t) = 3.67 - 0.37 \log_{10} (M_{\rm can}).
\label{eq:deltatMdyn}
\end{equation}
Note the difference between $\Delta t$ in the above equation and $\delta t$ in equation~(\ref{eq:MSFR}): $\delta t$ is the timescale that it takes a galaxy to form a representative population of star clusters that fully sample the IGIMF, while $\Delta t$ is an estimate for the total time that it takes the galaxy most of its stars. Thus, $\Delta t$ is typically of the order of several Gyr, while $\delta t$ is much shorter.

Equation~(\ref{eq:deltatMdyn}) is based on the canonical IMF, which is a deliberate choice, since also the initial values for the galaxy masses are based on the canonical IMF. However, for the final values in mass according to the IGIMF model, the formation timescales are likely more similar to equation~(19) in \citet{Recchi2009}, which has the same structure as equation~(\ref{eq:deltatMdyn}), but the coefficients have been adopted to the IGIMF. Above a mass of $\approx 10^8 - 10^9 \, {\rm M}_{\odot}$ though, the difference between the two equations seems negligible. Equation~(\ref{eq:deltatMdyn}) is a bit steeper than equation~(19) in \citet{Recchi2009} also for $M \apprge 10^8 \, {\rm M}_{\odot}$, but the two equations intersect at $M \approx 10^{10} \, {\rm M}_{\odot}$. Only below a mass of $10^8 \, {\rm M}_{\odot}$, the formation time scales differ more noticeably (but see the end of this Section why these values should be taken with care anyway). The same could also be said for masses higher than $10^{11} - 10^{12} \, {\rm M}_{\odot}$, but this is the upper end for galaxy formation and there are hardly any galaxies to test this. 

It may be a fortunate coincidence that equation~(\ref{eq:deltatMdyn}) from \citet{Thomas2005} and equation~(19) in \citet{Recchi2009} are nearly identical in much of the critical mass range. However, the upper and medium panels of figure~(7) in \citet{Fontanot2017}, which show the mean star formation histories and the mean mass assemblies of galaxies with masses between $10^{9.25} \, {\rm M}_{\odot}$ and $10^{12} \, {\rm M}_{\odot}$, confirm that the star formation time scale changes only little when the canonical IMF is changed for the IGIMF. The same is not true for the [O/{\rm Fe}]-ratio though, as the lower panels of figure~(7) in \citet{Fontanot2017} shows for the same galaxies. Thus, metallicities may change a lot when the star formation time scale changes hardly. However, in this paper, we are mainly interested in the star formation time scales, and not in metallicities. 

\citet{Thomas2005} used Gaussian functions with a width depending on the mass of the ETG for the formation time scale of the ETGs. The values for $\Delta t$ in equation~(\ref{eq:deltatMdyn}) are therefore variances of the Gaussian functions, or equivalently, the timescale it takes an ETG at the peak of its star formation to form 66 percent of its total stellar population. However, the most straight-forward way to link the $M_{\rm can}$ of the ETGs from \citet{Dabringhausen2016a} to SFRs, and thereby to values for $\alpha_{\rm eff}$ (cf. equations~\ref{eq:alphaeff21} to~\ref{eq:alphaeff23}), is to simply divide $M_{\rm can}$ of each ETG by the value of $\Delta t$ calculated for it with the equation~(\ref{eq:deltatMdyn}). This is the approach chosen here.

A fit to the data on $\alpha_{\rm eff}$ versus $M_{\rm can}$ holds
\begin{align}
\label{eq:Mdyn-alpha1}
\nonumber \alpha_{\rm eff} = & \frac{a_{SFR} \, \log_{10}(M_{can})-b_{SFR}}{1 -\exp[-c_{SFR} (a_{SFR} \, \log_{10}(M_{can})-b_{SFR})]} \\
+ & d_{SFR} \, \log_{10}(M_{can})+e_{SFR},
\end{align}
where $M_{\rm can}$ is in Solar units and $a_{\alpha}=-1.35$, $b_{\alpha}=-8.66$, $d_{\alpha}=-0.326$ and $e_{\alpha}=-5.42$ are parameters that are obtained in a least-square fit. The parameter $c_{SFR}$ is set to a fixed value of 10 here to ensure the convergence of the fit. Asymptotically, functions like the one expressed in equation~(\ref{eq:Mdyn-alpha1}) approach a linear function $\alpha_{\rm eff}=d \, \log_{10}(M_{\rm can})+e$ for low $M_{\rm can}$ and a linear function $\alpha_{\rm eff}=(a+d)\, \log_{10}(M_{\rm can}) + (b+e)$ for high $M_{\rm can}$. The larger $c$ is, the narrower is the transition from the low-$M_{\rm can}$-range to the high-$M_{\rm can}$ range in equation~(\ref{eq:Mdyn-alpha1}). This is relevant because the parameter $c$ is left free in some cases in equations~(\ref{eq:alphaefffit}) to~(\ref{eq:MLratiofit}), which have a similar structure. The data for the effective high-mass IGIMF-slope and the fit to them is shown in Figure~(\ref{fig:Mdyn-alpha}).

While this method for linking $\alpha_{\rm eff}$ to the star formation rates is simple, obections may come from that very fact. One objection would be that \citet{Thomas2005} considered the variances of Gaussian functions as a measure for the star formation time scales, whereas here, the variance of the Gaussian function is taken for the timescale it takes for the complete stellar population to form. Also, the Gaussian function changes, whereas here, an average value is taken for the star formation time scale. However, these values star formation time scales are only indicative as they do not consider the history of individual galaxies; for instance if they have experienced starbursts through mergers with other galaxies. Thus, the relation between $M_{\rm can}$ and the SFR is hardly an exact one, so that two ETGs with the same mass might have formed the bulk of their stars on somewhat different time scales. Also note that the assumption of a Gaussian function is itself only an approximation, even though a very good one according to \citet{Thomas2005}. Another objection would be that even the smallest galaxies are treated as SSPs, even though they have in reality often epochs of significant star formation almost over the age of the Universe (see \citealt{Weisz2014}). Again those values are only indicative for masses $M \apprle 10^9 \, {\rm M}_{\odot}$, where $M \approx 10^9 \, {\rm M}_{\odot}$ is the mass where $\alpha_{\rm eff} \approx 2.3$ for all passbands $L_{pb}$ considered here. $M \approx 10^9 \, {\rm M}_{\odot}$ corresponds to a period of star formation of approximately 1~Gyr with the assumptions made here. For masses $M < 10^9$, $\alpha_{\rm eff}$ increases above 2.3, but even for the most extreme case, $\alpha_{\rm eff}=\infty$, the mass of the galaxy is appoximately 0.7 times the mass of the same galaxy with the canonical IMF ($\alpha_{\rm eff}=2.3$), as can be seen Figs.~(\ref{fig:IGIMF_LV}) and~(\ref{fig:IGIMF_Lr}). In other words, the masses for small galaxies change only little, even for big changes in $\alpha_{\rm eff}$ (compare Figs.~\ref{fig:IGIMF_LV} and~\ref{fig:IGIMF_Lr} to Figs.~\ref{fig:IGIMF-alphaV} and~\ref{fig:IGIMF-alphar}). This is not true for galaxies with masses $M \apprge 10^9 \, {\rm M}_{\odot}$. For them, the star formation timescale is $\apprle$ 1~Gyr, which is generally much shorter than the age of the galaxies. Thus, the stellar populations approach SSPs at least in age. In this range, also the mass estimates become increasingly dependent on $\alpha_{\rm eff}$ (compare Fig.~\ref{fig:IGIMF-alphaV} to Fig.~\ref{fig:IGIMF_LV} and Fig.~\ref{fig:IGIMF-alphar} to Fig.~\ref{fig:IGIMF_Lr})

However, galaxy mergers are indeed assumed to make matters more complicated than described here. If two galaxies with ongoing star formation collide, the collision is thought to provoke a starburst, which is at least qualitatively consistent with the notion that more massive galaxies form stars more rapidly than lighter galaxies. However, imagine for instance two galaxies that merge after each of them has finished its star formation. According to the picture here, their formation time scales were those of the lighter progenitors, but its mass would be that of the more massive merger remnant. In the end, the time scales for galaxy formtion cited here are just statistical numbers that indicate how an average ETG of a certain mass is supposed to evolve, while individual ETGs of that mass can deviate from this value quite strongly.

\subsection{The masses of ETGs according to the IGIMF-model}
\label{sec:masses-ETGs}

\begin{figure*}p
\centering
\includegraphics[scale=0.85]{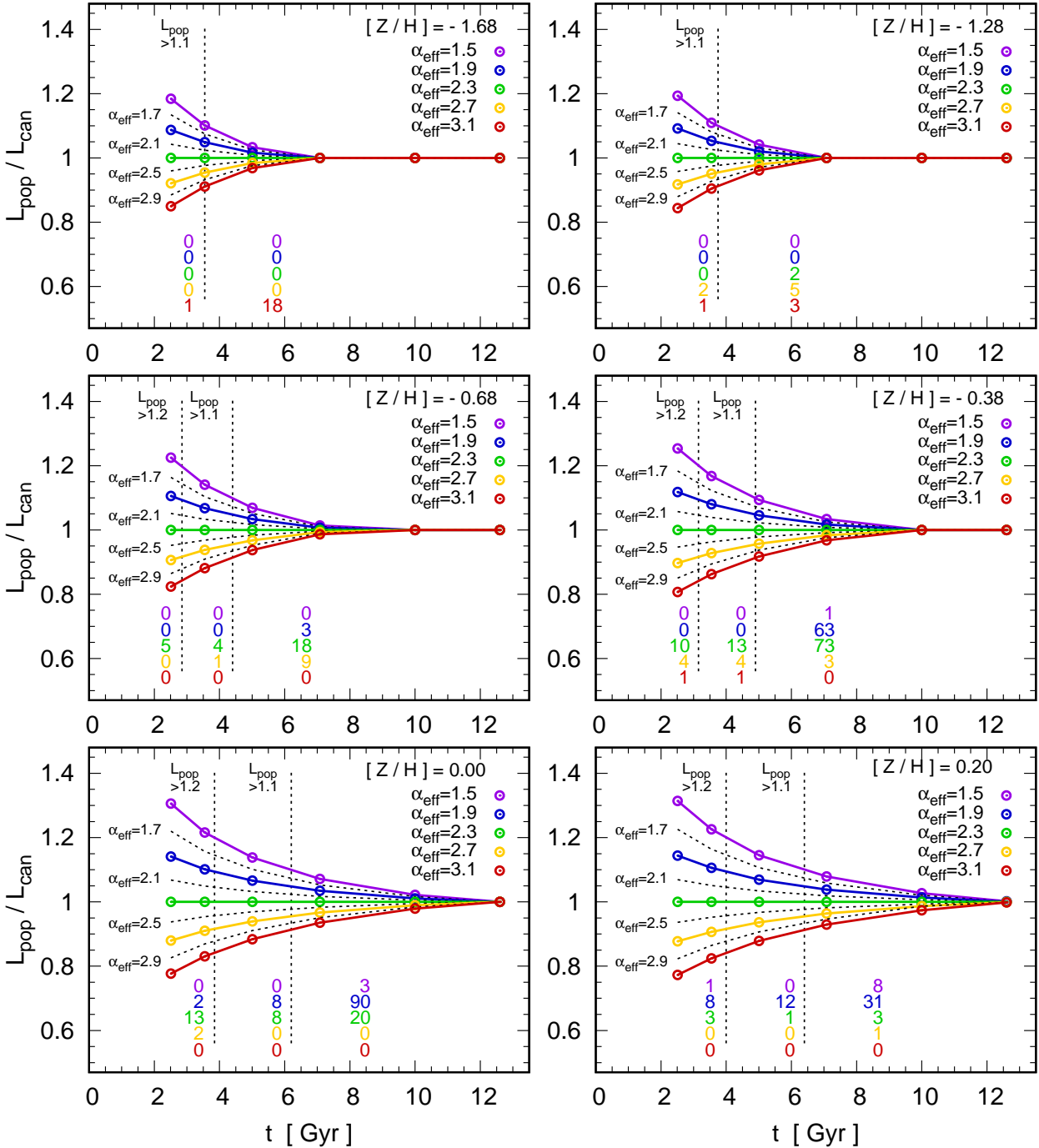}
\caption[The luminosity of stellar populations with different IGIMFs over age for different metallicities]{\label{fig:Lratios-Z} The luminosity of stellar populations with different IGIMFs over age for different metallicities. The coloured dots indicate the luminosities for specific choices for the metallicity ([Z/H]), the ages ($t$) and the high-mass slopes ($\alpha_{\rm eff}$). Each panel is dedicated to one metallicity that is given in the upper right corner of that panel, and the considered ages are 2.51 Gyr, 3.55 Gyr, 5.01 Gyr, 7.08 Gyr, 10.0 Gyr and 12.6 Gyr. The different high-mass slopes are encoded with different colours: violet for $\alpha_{\rm eff}=1.5$, blue for $\alpha_{\rm eff}=1.9$, green for $\alpha_{\rm eff}=2.3$ (i.e. the high-mass slope of the canonical IMF), yellow for $\alpha_{\rm eff}=2.7$ and red for $\alpha_{\rm eff}=3.1$. The lines connecting the dots approximate the evolution of a stellar population with the respective high-mass slope over time. They converge into a single (red) line for ages where $m_{\rm to} \le 1 {\rm M}_{\odot}$. The dashed lines define a grid in each panel. The vertical dashed lines indicate where $L_{\rm pop}(\alpha_{\rm eff})/L_{\rm pop}(\alpha_{\rm eff}=2.3)$ can surpass 1.1, and 1.2 respectively, for a very top-heavy IGIMF. The intersecting dashed lines indicate the time evolution of $L_{\rm pop}(\alpha_{\rm eff})/L_{\rm pop}(\alpha_{\rm eff}=2.3)$ for $\alpha_{\rm eff}=1.7$, $\alpha_{\rm eff}=2.1$, $\alpha_{\rm eff}=2.5$ and $\alpha_{\rm eff}=2.9$ from top to bottom. The number of ETGs is counted in each grid cell. Note that only ETGs with known spectroscopic ages are considered in this plot, and the metallicities of the ETGs are in intervals that are centered on the value quoted in the upper right corners of each panel. The numbers of galaxies in the grid cells are noted below the grid in two or three columns (depending on the number of vertical dashed lines in the respective panel), and coded with the colour of the evolutionary line that runs through the addressed grid cell.}
\end{figure*}

Estimates for the stellar mass of an ETG according to the IGIMF-model, $M_{\rm IGIMF}$, can be obtained based on existing estimates for $M_{\rm can}$, i.e. mass estimates that are based on equation~(\ref{eq:IMF}) with $m_{\rm max}= 100 \ {\rm M}_{\odot}$, and estimates for the masses of evolved SSPs that formed with the IGIMF. The latter are given as
\begin{equation}
M_{\rm pop}(\alpha_{\rm eff})=k_{\rm L}\int^{m_{\rm max}}_{0.1} m_{\rm rem}(m) \xi_{\rm IGIMF}(m) \, dm,
\label{eqMrem2}
\end{equation}
i.e. by integrating the present-day masses of stars and stellar remants over the whole range of initial stellar masses. In equation~(\ref{eqMrem2}), $\xi_{\rm IGIMF}(m)$ is the IGIMF given by equation~(\ref{eq:IGIMF3}), $m_{\rm max}$ is the mass of the most massive stars forming in the galaxy and is given by equation~(\ref{eq:interpolationMmax}) for $\log_{10}(SFR/{\rm M}_{\odot}{\rm yr}^{-1})<-0.878$ or $m_{\rm max}=150 \ {\rm M}_{\odot}$ for $\log_{10}(SFR/{\rm M}_{\odot}{\rm yr}^{-1}) \ge -0.878$, and $m_{\rm rem}(m)$ is the initial-to-final mass function, which expresses the masses of stellar remnants as a function of the initial mass. The factor $k_{\rm L}$ is a scaling factor that ensures that the total luminosity of the stellar population remains constant when $\alpha_{\rm eff}$ varies. This is motivated by the notion that the luminosity is fixed through observations, while the mass of the stellar population is treated here as an unknown parameter that is to be determined.

The initial-to-final mass function used in equation~(\ref{eqMrem2}) is the one introduced in \citet{Dabringhausen2009}, which is designed for stellar systems where star formation has ceased at least of the order of $10^8$ years ago. It is given as
\begin{equation}
\label{eq:mrem}
m_{\rm rem}=
\begin{cases}
\frac{m}{{\rm M}_{\odot}} & {\rm if \ \ } \frac{m}{{\rm M}_{\odot}} < \frac{m_{\rm to}}{{\rm M}_{\odot}}\\
0.109 \, \frac{m}{{\rm M}_{\odot}}+0.394 & {\rm if \ \ } \frac{m_{\rm to}}{{\rm M}_{\odot}} \le \frac{m}{{\rm M}_{\odot}} < 8\\
1.35 & {\rm if \ \ } 8  \le  \frac{m}{{\rm M}_{\odot}} < 25\\
const \, \frac{m}{{\rm M}_{\odot}} & {\rm if \ \ } \leq \frac{m}{{\rm M}_{\odot}} \le m_{{\rm max}}.
\end{cases}
\end{equation}
{$m_{\rm max}$ is given in this equation by equation~(\ref{eq:interpolationMmax}) for $\log_{10} (SFR/{\rm M}_{\odot}{\rm yr}^{-1})< -0.878$ and by $150 \, {\rm M}_{\odot}$ else.}

Thus, stars with masses below the main-sequence turn-off mass $m_{\rm to}$ are considered to still have their initial masses, white dwarfs are thought to have progenitors with masses between $m_{\rm to}$ and $8 \, {\rm M}_{\odot}$ and their masses are given by a relation found by \citet{Kalirai2008} and neutron stars are thought to have progenitors with masses between $8 \, {\rm M}_{\odot}$ and $25 \, {\rm M}_{\odot}$ and are all considered to have a mass of $1.35 \, {\rm M}_{\odot}$, which is observationally supported by \citet{Thorsett1999}. Stars with initial masses above $25 \, {\rm M}_{\odot}$ are considered to evolve into black holes. The mass of these black holes is the most uncertain parameter and strongly depends on their metallicity (compare for example figures~12 and~16 in \citealt{Woosley2002}). In equation \ref{eq:mrem}, the black hole mass is set to a constant fraction between 0 and 1 of the progenitor stars. The existence of black holes with masses well above 10 ${\rm M}_{\odot}$ was recently confirmed observationally via the detection of gravitational waves \citep{Abbott2016}, while less massive black holes have already been found before via the X-ray radiation that such a black hole emits if it accretes matter.

In theory, the case of low black hole masses corresponds to stars with high metallicities, and the case with high black hole masses to stars with low metallicities. This progression of the mass of the black holes with metallicity is estimated here as
\begin{equation}
\label{blackhole}
m_{\rm BH}=\left(-\frac{4}{3} [Z/{\rm H}] + \frac{1}{6}\right) m_{\rm star},
\end{equation}
where $m_{\rm BH}$ is the mass of the black hole, $[Z/{\rm H}]$ is the metallicity and $m_{\rm star}$ is the mass of the progenitor star. Equation~\ref{blackhole} produces black holes that have 0.5 times the mass of the progenitor at $[Z/{\rm H}]=-2.5$ and decline linearly to 0.1 times the progenitor mass at $[Z/{\rm H}]=0.5$. The explicit terms that the integration in equation~(\ref{eqMrem2}) yields are listed in the appendix to \citet{Dabringhausen2009}, provided that 0.1 in their last equation is replaced by $-4/3[Z/{\rm H}]+1/6$.

The estimates for the actual masses of the stellar populations of the ETGs according to the IGIMF-model can then be calculated as
\begin{equation}
M_{\rm IGIMF}=\left(\frac{M_{\rm pop}(\alpha_{\rm eff})}{M_{\rm pop}(\alpha_{\rm eff}=2.3)}\right)\times M_{\rm can},
\label{eq:MIGIMF}
\end{equation}
where $M_{\rm pop}(\alpha_{\rm eff})$ is given by evaluating equation~(\ref{eqMrem2}) for the $\alpha_{\rm eff}$ implied by the SFR with which the majority of the stellar population of the studied ETG has formed and $m_{\rm max}$ is given by equation~(\ref{eq:interpolationMmax}) for a maximum stellar mass of $150 \, {\rm M}_{\odot}$, $M_{\rm pop}(\alpha_{\rm eff}=2.3)$ is given by evaluating equation~(\ref{eqMrem2}) for $\alpha_{\rm eff}=2.3$ and $m_{\rm max}=150 \ {\rm M}_{\odot}$, and $M_{\rm can}$ is the estimate of the mass that the stellar population of the studied ETG would have with $\alpha_{\rm eff}=2.3$ and $m_{\rm max}=100 \ {\rm M}_{\odot}$. The reason why in the estimates for $M_{\rm can}$ an upper mass limit of $m_{\rm max}=100 \ {\rm M}_{\odot}$ instead of $m_{\rm max}=150 \ {\rm M}_{\odot}$ is assumed is that $m_{\rm max}=100 \ {\rm M}_{\odot}$ is also the upper mass limit in the SSP-models by \citet{Bruzual2003}. The models by \citet{Bruzual2003} are the basis for the models discussed here, but $150 \, {\rm M}_{\odot}$ is the more realistic choice, which becomes relevant for top-heavy IGIMFs. However, the canonical IMF is steep enough at high stellar masses that the difference between $m_{\rm max}=100 \ {\rm M}_{\odot}$ and $m_{\rm max}=150 \ {\rm M}_{\odot}$ is rather inconsequential for $M_{\rm can}$.

Assuming that the ETGs contain at least in their inner regions (up to a few half-light radii) no significant amounts of non-baryonic dark matter, $M_{\rm IGIMF}$ should be approximately equal to the dynamical mass of the ETGs, $M_{\rm dyn}$, since ETGs generally contain also little gas \citep{Young2011} and dust \citep{Dariush2016}. If it is found that $M_{\rm IGIMF}<M_{\rm dyn}$ on average for the ETGs, this can conversely be interpreted as unaccounted matter, which may be either non-baryonic or provided by an IGIMF that is systematically different than assumed. However, finding $M_{\rm IGIMF}>M_{\rm dyn}$ on average would unambiguosly exclude the assumed IGIMF, assuming that $M_{\rm dyn}$ is an estimate for the total mass of a stellar system in virial equilibrium.

The estimate of $M_{\rm IGIMF}$ with equations~(\ref{eqMrem2}) to~(\ref{eq:MIGIMF}) is simplified a lot by the observation that the stellar populations in ETGs are typically signficantly older than 1 Gyr. This implies that the most massive stars that have not evolved into essentially non-luminous remnants yet have masses of about 1 ${\rm M}_{\odot}$. This is because the turn-off mass, ${m_{\rm to}}$, of a young stellar population very quickly approaches 1 ${\rm M}_{\odot}$ as the stellar population ages. For instance, for a 2.5 Gyr old population, ${m_{\rm to}}$ is according to the stellar models by \citet{Girardi2000} already between 1.32 ${\rm M}_{\odot}$ and 1.61 ${\rm M}_{\odot}$, depending on the metallicity, and there is hardly a Galaxy with a lower spectroscopic age in the sample considered here. These values change only little as the stellar population changes further, so that ${m_{\rm to}}$ lies between 0.88 ${\rm M}_{\odot}$ and 1.07 ${\rm M}_{\odot}$ after 10 Gyr. The IGIMF is moreover formulated here (and similarly in other papers) such that it does not vary below 1 ${\rm M}_{\odot}$, and the IGIMFs in galaxies with different SFRs begin to diverge from each other only for masses $m > 1 \ {\rm M}_{\odot}$\footnote{Note that this formulation of the IGIMF also implies that a galaxy always produces low-mass stars, independent of whether it produces also high-mass stars or not. This is desirable, since plenty of low-mass stars are generally observed in galaxies.}.

These properties of the IGIMF and the proximity of $m_{\rm to}$ to 1 ${\rm M}_{\odot}$ for all stellar populations with ages between a few Gyr and a Hubble time motivate to set the parameter $m_{\rm to}$ in equation~(\ref{eq:mrem}) to 1 ${\rm M}_{\odot}$ for simplicity. As a consequence of this approximation, the normalisation factor $k_{\rm L}$ in equation~(\ref{eqMrem2}) is the same for all IGIMFs considered in this paper, since it was introduced only to ensure that the stellar population of a galaxy always produces its observed luminosity, independent of a change of the parameters that determine the IGIMF. It is therefore sufficient to evaluate only the integral in equation~(\ref{eqMrem2}) and ignore $k_L$, since $k_L$ is a constant factor that cancels out in equation~(\ref{eq:MIGIMF}).

To test if the approximation of setting $m_{\rm to}$ to 1 for all stellar populations can be justified in this paper, the luminosity of stellar populations is calculated in dependency of the shape of the IGIMF. For this purpose, a grid of stellar isochrones with different ages and metallicities from \citet{Girardi2000} is used, as well as the the IGIMF as formulated in equation~(\ref{eq:IGIMF3}), but without a forefactor that normalises it to any praticular luminosity or mass. Not normalising equation~(\ref{eq:IGIMF3}) emulates the assumption that $m_{\rm to}=1 \ {\rm M}_{\odot}$ in all considered galaxies, which implies that the normalisation factor $k_L$ in equation~(\ref{eqMrem2}) is a constant. More explicitly, this luminosity is calculated by numerically integrating
\begin{equation}
\label{eq:LIGIMF}
L_{\rm pop}(\alpha_{\rm eff})=\int_{0.1}^{m_{\rm max}} l(m) \xi_{\rm IGIMF}(m, \alpha_{\rm eff}) dm,
\end{equation}
for different $\alpha_{\rm eff}$, where $\xi_{\rm IGIMF}$ describes the shape of the IGIMF as a function of stellar mass as given in equation~(\ref{eq:IGIMF3}), $\alpha_{\rm eff}$ is the high-mass slope of the IGIMF, $l(m)$ is the luminosity density as a function of stellar mass, and $m_{\rm max}$ is the upper mass limit for stars. All masses and luminosities in equation~(\ref{eq:LIGIMF}) are in Solar units.

The luminosities of the integrated stellar populations for different metallicities, ages and $\alpha_{\rm eff}$ are shown over age in Fig.~(\ref{fig:Lratios-Z}). The results for $L_{\rm pop}(\alpha_{\rm eff})$ are divided by $L_{\rm pop}(\alpha_{\rm eff}=2.3)$. In other words, they are given in units of the luminosity of the canonical IMF, provided that the canonical IMF has the same normalisation like the IGIMF, and IMF and IGIMF are therefore equal below 1 ${\rm M}_{\odot}$. The coloured lines in Fig.~(\ref{fig:Lratios-Z}) indicate the evolution of $L_{\rm pop}(\alpha_{\rm eff})/L_{\rm pop}(\alpha_{\rm eff}=2.3)$ with age. At certain metallicity-dependent ages, the lines merge into a single line with $L_{\rm pop}(\alpha_{\rm eff})/L_{\rm pop}(\alpha_{\rm eff}=2.3)=1$. At and above this age, $m_{\rm to} \le 1  \ {\rm M}_{\odot}$ is fulfilled. Below this age, the lines diverge and indicate by how much $L_{\rm pop}(\alpha_{\rm eff})$ of a stellar population with a given IGIMF deviates from $L_{\rm pop}(\alpha_{\rm eff}=2.3)$, or equivalently from the luminosity of the canonical IMF. The lines show that only stellar populations with fairly young SSP-equivalent ages ($t \apprle$ 4 Gyr) and extreme IGIMFs ($\alpha_{\rm eff} \apprle 1.7$ or $\alpha_{\rm eff} \apprge 3.0$) are expected to have luminosities that deviate by 20 per cent or more from the prediction from the canonical IMF. This is still fairly moderate compared to the deviations in mass calculated with equation~(\ref{eq:MIGIMF}) and discussed in the next sections, but another question is how many ETGs in the sample discussed here actually fall into these extreme categories.

To answer this question, grids are placed over the panels of Figure~(\ref{fig:Lratios-Z}). The vertical dashed lines indicate where $L_{\rm pop}(\alpha_{\rm eff})/L_{\rm pop}(\alpha_{\rm eff}=2.3)$ can for a very top-heavy IMF surpass 1.1, and 1.2 respectively. The intersecting dashed lines indicate the time evolution of $L_{\rm pop}(\alpha_{\rm eff})/L_{\rm pop}(\alpha_{\rm eff}=2.3)$ for $\alpha_{\rm eff}=1.7$, $\alpha_{\rm eff}=2.1$, $\alpha_{\rm eff}=2.5$ and $\alpha_{\rm eff}=2.9$ from top to bottom. The ETGs in the sample considered here are grouped in six metallicity intervals, namely $[Z/H]<-1.48$, $-1.48 < [Z/H] \le -0.98$, $-0.98 < [Z/H] \le -0.53$, $-0.53 < [Z/H] \le -0.19$, $-0.19 < [Z/H] \le 0.10$ and $[Z/H] > 0.10$. Thus, the intervals are defined such that they are centered on the metallicities given in the upper right corners of Figure~(\ref{fig:Lratios-Z}), which are the metallicities of the stellar isochrones from \citet{Girardi2000} that were used to calculate $L_{\rm pop}(\alpha_{\rm eff})$. To each panel is thereby a group of ETGs assigned. The number of ETGs in each grid cell can then be counted.

The numbers resulting from the counts in the grid cells are noted in the lower half of each panel. They are coded with the colour of the evolutionary line that runs through the according grid cell. For instance, the blue number in the right column in the bottom left panel means that there are 90 ETGs with near-Solar metallicity, known spectroscopic ages $t>6.2$ Gyr and predicted high-mass slopes $1.7 \le \alpha_{\rm eff} < 2.1$ in that cell. Note that this combination of parameters implies $L_{\rm pop}(\alpha_{\rm eff})/L_{\rm pop}(\alpha_{\rm eff}=2.3)<1.1$ for all of these 90 ETGs.

Comparing the numbers in the grid cells, it turns out that objects young enough to fulfill $L_{\rm pop}(\alpha_{\rm eff})/L_{\rm pop}(\alpha_{\rm eff}=2.3)>1.1$ at least for very high or very low values of $\alpha_{\rm eff}$ are quite rare, namely 101 out of the total of 463 ETGs. For objects young enough that $L_{\rm pop}(\alpha_{\rm eff})/L_{\rm pop}(\alpha_{\rm eff}=2.3)>1.2$ can be fulfilled in principle, the number diminishes to 53 out of 463 ETGs. Further restricting the sample to objects that additonally fulfill $\alpha_{\rm eff} < 1.7$ or $\alpha_{\rm eff} > 2.9$ limits it to 5 ETGs for $L_{\rm pop}(\alpha_{\rm eff})/L_{\rm pop}(\alpha_{\rm eff}=2.3)>1.1$, and 2 ETGS for $L_{\rm pop}(\alpha_{\rm eff})/L_{\rm pop}(\alpha_{\rm eff}=2.3)>1.2$, respectively. Thus, objects that due to a combination of low age and an extreme IGIMF have a significant deviation from the luminosity predicted for the canonical IMF (say more than 10 per cent) are very rare. Assuming $m_{\rm to}=1 \ {\rm M}_{\odot}$, and thus $L_{\rm pop}(\alpha_{\rm eff})/L_{\rm pop}(\alpha_{\rm eff}=2.3)=1$ for all ETGs therefore seems a rather good approximation for the purpose of this paper.

\section[Results]{Results}
\label{sec:results}

\begin{figure}
\centering
\includegraphics[scale=0.85]{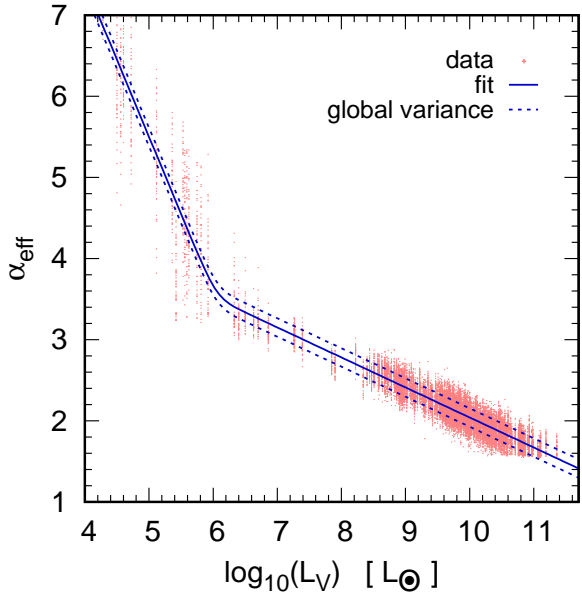}
\caption{\label{fig:IGIMF-alphaV} The effective high-mass slope of the IGIMF, $\alpha_{\rm eff}$, in dependency of the $V$-band luminosity, $L_V$. Each ETG is represented 50 times with different masses that are based on the mass given in \citet{Dabringhausen2016a}, but have been modified by adding random representations of a Gaussian distribution with a width of $\sigma=0.3$. The drawn line is a fit of equation~(\ref{eq:alphaefffit}) to the data. The dashed lines are same as the drawn line, except that constant values have been substracted or added to it. The range between the two dashed lines is the range in which the central 2/3 of the data lie.}
\end{figure}

\begin{figure}
\centering
\includegraphics[scale=0.85]{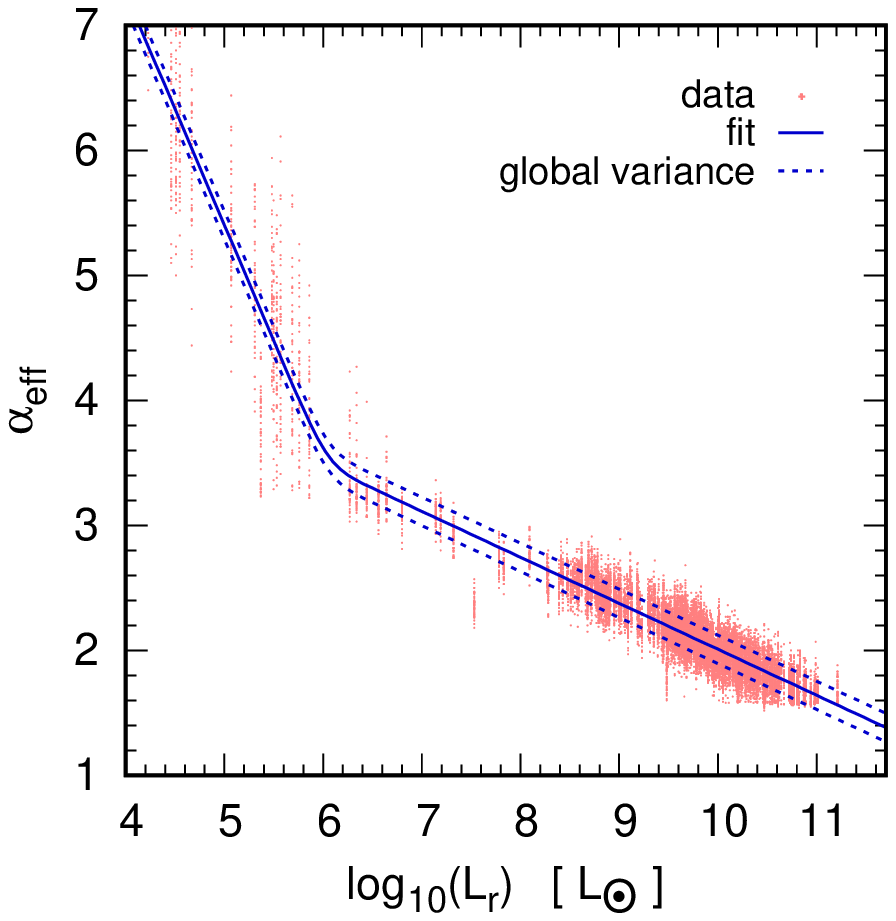}
\caption{\label{fig:IGIMF-alphar} As figure~\ref{fig:IGIMF-alphaV}, but for the $r$-band instead of the $V$-band.}
\end{figure}

\begin{figure}
\centering
\includegraphics[scale=0.85]{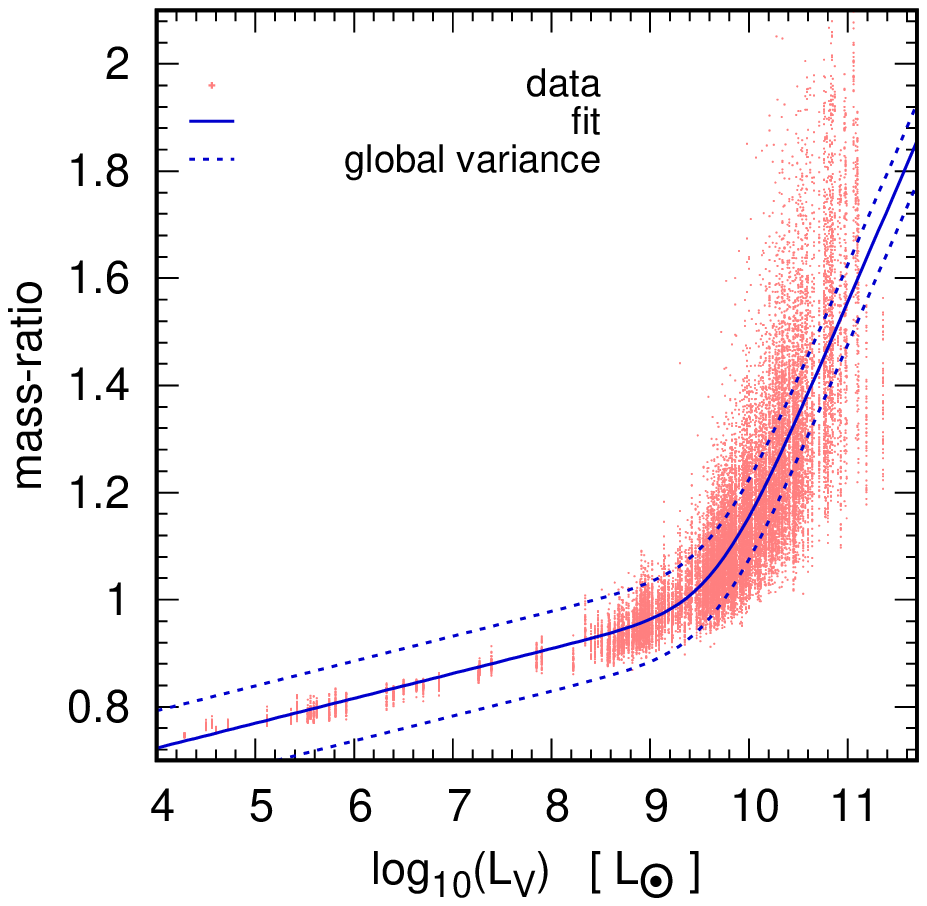}
\caption[The effective high-mass IGIMF-slope and the ratio between the stellar mass according to the canonical IMF and the the stellar mass according to the IGIMF in dependency of the $V$-band luminosity]{\label{fig:IGIMF_LV} The mass-ratio between the IGIMF and the canonical IMF of the ETGs, $M_{\rm IGIMF}/M_{\rm can}$, in dependency of the $V$-band luminosity, $L_V$. Each ETG is represented 50 times with different masses that are based on the mass given in \citet{Dabringhausen2016a}, but have been modified by adding random representations of a Gaussian distribution with a width of $\sigma=0.3$. The drawn line is a fit of equation~(\ref{eq:Mratiofit}) to the data. The dashed lines are same as the drawn line, except that constant values have been substracted or added to it. The range between the two dashed lines is the range in which the central 2/3 of the data lie.}
\end{figure}

\begin{figure}
\centering
\includegraphics[scale=0.85]{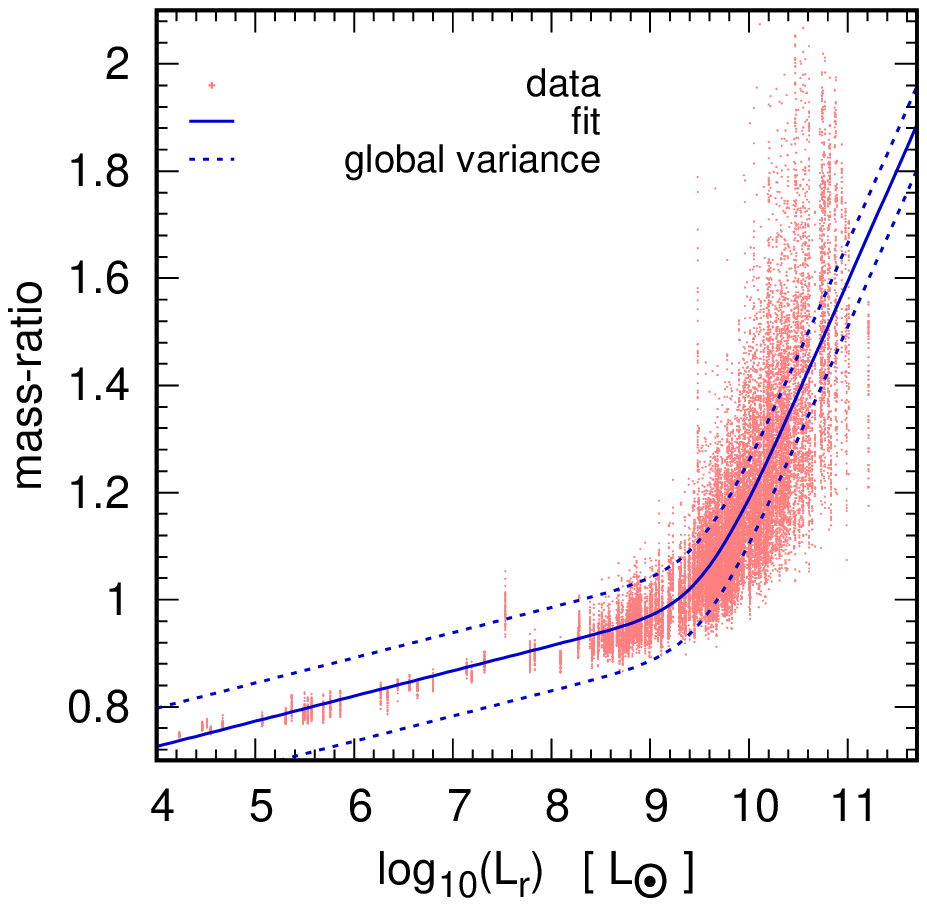}
\caption[The effective high-mass IGIMF-slope and the ratio between the stellar mass according to the canonical IMF and the the stellar mass according to the IGIMF in dependency of the $r$-band luminosity]{\label{fig:IGIMF_Lr} As figure~(\ref{fig:IGIMF_LV}), but for the $r$-band instead of the $V$-band.}
\end{figure}

Having assigned values for $\alpha_{\rm eff}$ to the galaxies taken from \citet{Dabringhausen2016a}, interpolation relations for the dependency of $\alpha_{\rm eff}$ and the stellar mass implied by $\alpha_{\rm eff}$ on luminosity can be established. The considered luminosities are the ones considered in the catalogue by \citet{Dabringhausen2016a} (i.e. the $U$, $B$, $V$ and $I_C$-passbands in the standard Johnson-Cousins magnitude system and the $u$, $g$, $r$, $i$ and $z$-passbands in the SDSS magnitude system). Also the mass according to the canonical IMF is considered, since the IGIMF is in most cases different from the canonical IMF, which is the basis for the estimates of the IGIMF (see Section~\ref{sec:data-masses}).

The mass the galaxies is a rather uncertain issue. The first factor for that uncertainty in mass mentioned here is the treatment of the ETGs as SSPs instead of multiple stellar populations (see Section~\ref{sec:data-masses}). A comparison between the two mass estimates for the galaxies in the ATLAS$^{3D}$ sample shows that the overall deviance can be described roughly with a gaussian with a variance of $\sigma=0.3$ (see figure~\ref{fig:SSP-comp}). While this imprecision is estimated for a specific set of galaxies, it affects galaxies in general, even though numerically possibly different. The second factor mentioned here is the estimate of a magnitude from other magitude(s) (Section~\ref{sec:data-luminosities}), which is in the end also transformed in an estimate of the mass. This uncertainty can be modelled by a Gaussian with a variance of 0.2 at most, while the number of galaxies that would have to be treated that way depends on the passband (e.g. less than 10 per cent in the $B$-band and a bit more than one half in the $U$-band; see  Section~\ref{sec:data-luminosities}). Taking $\Delta=\sqrt{\Delta_1+\Delta_2+ \dots}$ for the error propagation, where $\Delta$ is the total error and $\Delta_i$ are its components, it becomes clear that the error is dominated to a large degree by the uncertainty due to the treatment of the stellar populations as SSPs. Thus, the mass of each galaxy is represented by 50 entries of the form $\log_{10}(M_{i,j})=log_{10}(M_i)+\Delta_j$, where $log_{10}(M_i)$ is the logarithmic mass of the $i$th galaxy according to \citet{Dabringhausen2016a} and the $\Delta_j$ is a number randomly drawn from a gaussian with a variance of $0.3$.

While taking the uncertainty due to mass into account does make some sense, dividing the sample of the 460 galaxies into subsamples according to their source would not. The reason is that each subsample is directed at galaxies in comparatively narrow mass range; e.g. \citealt{McConnachie2012} at small Local Group ETGs, \citealt{Guerou2015} for somewhat more massive ETGs, and the ATLAS$^{3D}$ sample \citep{Cappellari2011} for the most massive ETGs. Naturally, a fit to the data of a subsample runs through the sample itself, but has no predictive power below or above the range of the subsample. Thus, functions which descibe the run of the data over the whole mass range can only be derived from the sample as a whole. What gives the results credence is either the work of other groups who arrive at similar results for ETGs in the same mass range, or the continuity of the results with results in neighbouring samples. The subsamples here fulfill both criteria, except the small-scale ETGs from the Local Group by \citet{McConnachie2012}, which are merely the low-mass continuation of the more massive ETGs.

Thus, the effective high-mass IGIMF slope of the sample as a whole can be expressed as
\begin{equation}
\label{eq:alphaefffit}
\alpha_{\rm eff}=\delta_{\alpha}\left(\frac{a_{\alpha} \, x-b_{\alpha}}{1-\exp[-c_{\alpha} (a_{\alpha} \, x-b_{\alpha})]}\right)+d_{\alpha} \, x+e_{\alpha}
\end{equation}
where $x$ is the base-10 logarithm of a luminosity or the stellar mass under the assumption that the IMF is canonical (taken from \citealt{Dabringhausen2016a}) and $a_{\alpha}$, $b_{\alpha}$, $c_{\alpha}$, $d_{\alpha}$ and $e_{\alpha}$ are parameters that are obtained in least-square fit and are listed in table~(\ref{tab:alphaefffit}).

The parameter $\delta_{\alpha}$ in equation~(\ref{eq:alphaefffit}) can assume the values 0 and 1. If $\delta_{\alpha} = 1$, equation~(\ref{eq:alphaefffit}) (and also the similar equations~\ref{eq:Mratiofit} and~\ref{eq:MLratiofit} further below) has the same basic structure like equation~(\ref{eq:Mdyn-alpha1}). Thus, it approximates a linear function that changes into a different linear function over a certain range in $x$. If $\delta_{\alpha} = 0$, the first term in equation~(\ref{eq:alphaefffit}) disappears and equation~(\ref{eq:alphaefffit}) becomes $\alpha_{\rm eff} = d_{\alpha} \, x+e_{\alpha}$.

\begin{landscape}
\begin{table}
\caption{\label{tab:alphaefffit} Best-fitting parameters in equation~(\ref{eq:alphaefffit}) for different passbands as well as $M_{\rm can}$, i.e the mass of the stellar population suggested by a SSP-model from \citet{Bruzual2003} with the canonical IMF. Column~1 specifies the variable considered in equation~(\ref{eq:alphaefffit}) and Column~2 gives the approximate range over which the function was fitted. Column~3 gives the number of galaxies used in the fit of the parameters. Columns~4,~5,~9,~11 and~13 list the obtained values for the parameters $\delta_{\alpha} $, $a_{\alpha}$, $b_{\alpha}$, $d_{\alpha}$ and $e_{\alpha}$, respectively. The numbers in brackets in columns~6,~8,~12 and~14 are the asymptotic standard errors in per cent to the according values. Thus, a value of more than 20 per cent shows that the asymptotic standard error is less than five $\sigma$ different from zero, provided the distribution is Gaussian. The $(*)$ in column~11 indicates that $c_{\alpha}$ in column~10 was not considered as a free parameter in the fit, but set to 10 in order to ensure convergence of the fitting routine. If a parameter is not specified because the whole term in which it appears is set to 0, the according entry in the table is set to $-/-$. Column~15 lists the value that to be substracted from equation~(\ref{eq:alphaefffit}) that globally 1/6 of the datapoints are below the function defined by the parameters in columns~4,~5,~7,~9,~11 and~13. Column~16 lists the value that to be added to equation~(\ref{eq:alphaefffit}) that globally 1/6 of the datapoints are above the function defined by the parameters in columns~4,~5,~7,~9,~11 and~13. Thus, Columns~15 and~16 define the range where the central 2/3 of the datapoints are located.}
\centering
\vspace{2mm}
\begin{tabular}{rrrrrrrrrrrrrrrr}
\hline
&&&&&&&&&&&&&&&\\[-10pt]
$x$      & range & $N$  & $\delta_{\alpha} $ & $a_{\alpha}$ & & $b_{\alpha}$ & & $c_{\alpha}$ & & $d_{\alpha}$ & & $e_{\alpha}$ & & $\Delta_-$ & $\Delta_+$ \\
\hline
$\log_{10}(L_u/{\rm L}_{\odot})$ & 3 to 10.5 & $ 460$ & $1$ & $-0.941$ & $(1.2\%)$ & $-6.115$ & $(1.0\%)$ & $10$ & $(*)$  & $-0.398$ & $(0.70\%)$  & $5.863$ & $(0.45\%)$ & $-0.114$  & $0.136$ \\
$\log_{10}(L_g/{\rm L}_{\odot})$ & 3 to 11   & $ 460$ & $1$ & $-0.861$ & $(1.2\%)$ & $-5.353$ & $(1.0\%)$ & $10$ & $(*)$  & $-0.376$ & $(0.65\%)$  & $5.733$ & $(0.41\%)$ & $-0.123$  & $0.126$ \\
$\log_{10}(L_r/{\rm L}_{\odot})$ & 3 to 11   & $ 460$ & $1$ & $-1.485$ & $(0.67\%)$& $-8.978$ & $(0.58\%)$& $10$ & $(*)$  & $-0.368$ & $(0.41\%)$  & $5.688$ & $(0.26\%)$ & $-0.117$  & $0.120$ \\
$\log_{10}(L_i/{\rm L}_{\odot})$ & 7 to 11   & $ 393$ & $0$ & $-/-$    &           & $-/-$    &           & $-/-$&        & $-0.353$ & $(0.45\%)$  & $5.568$ & $(0.28\%)$ & $-0.112$  & $0.121$ \\
$\log_{10}(L_z/{\rm L}_{\odot})$ & 7 to 11.5 & $ 347$ & $0$ & $-/-$    &           & $-/-$    &	          & $-/-$&        & $-0.370$ & $(0.42\%)$  & $5.787$ & $(0.27\%)$ & $-0.111$  & $0.114$ \\
$\log_{10}(L_U/{\rm L}_{\odot})$ & 7 to 11   & $ 429$ & $0$ & $-/-$    & 	   & $-/-$    &           & $-/-$&        & $-0.398$ & $(0.39\%)$  & $5.944$ & $(0.25\%)$ & $-0.119$  & $0.120$ \\
$\log_{10}(L_B/{\rm L}_{\odot})$ & 3 to 11   & $ 460$ & $1$ & $-0.843$ & $(1.2\%)$ & $-5.386$ & $(1.0\%)$ & $10$ & $(*)$  & $-0.381$ & $(0.67\%)$  & $5.819$ & $(0.42\%)$ & $-0.122$  & $0.124$ \\
$\log_{10}(L_V/{\rm L}_{\odot})$ & 3 to 11.5 & $ 460$ & $1$ & $-1.530$ & $(0.61\%)$& $-9.258$ & $(0.54\%)$& $10$ & $(*)$  & $-0.370$ & $(0.37\%)$  & $5.741$ & $(0.24\%)$ & $-0.114$  & $0.113$ \\
$\log_{10}(L_I/{\rm L}_{\odot})$ & 7 to 11   & $ 347$ & $0$ & $-/-$    &  	   & $-/-$    &           & $-/-$&        & $-0.390$ & $(0.40\%)$  & $5.942$ & $(0.26\%)$ & $-0.112$  & $0.115$ \\
$\log_{10}(M_{\rm can}/{\rm M}_{\odot})$ & 3 to 12 & $ 460$ & $1$ & $-1.364$ & $(0.031\%)$& $-8.710$ & $(0.026\%)$ & $10$ & $(*)$   & $-0.3245$   & $(0.020\%)$ & $5.402$ & $(0.012\%)$ & $-0.0041$ & $0.0034$ \\
&&&&&&&&&&&&&&&\\[-10pt]
\hline
\end{tabular}
\end{table}

\begin{table}
\caption{\label{tab:Mratiofit} Best-fitting parameters in equation~(\ref{eq:Mratiofit}) for different passbands as well as $M_{\rm can}$, i.e the mass of the stellar population suggested by a SSP-model from \citet{Bruzual2003} with the canonical IMF. Column~1 specifies the variable considered in equation~(\ref{eq:Mratiofit}) and Column~2 gives the approximate range over which the function was fitted. Column~3 gives the number of galaxies used in the fit of the parameters. Columns 4,~6,~8,~10 and~12 list the obtained values for the parameters $a_{mr}$, $b_{mr}$, $c_{mr}$, $d_{mr}$ and $e_{mr}$, respectively. The numbers in brackets in columns~5,~7,~9,~11 and~13 are the asymptotic standard errors in per cent to the values before them. Thus, a value of more than 20 per cent shows that the asymptotic standard error is less than five $\sigma$ different from zero, provided the distribution is Gaussian. Column~14 lists the value that to be substracted from equation~(\ref{eq:Mratiofit}) that globally 1/6 of the datapoints are below the function defined by the parameters in columns~4,~6,~8,~10 and~12. Column~15 lists the value that to be added to equation~(\ref{eq:Mratiofit}) that globally 1/6 of the datapoints are above the function defined by the parameters in columns~4,~6,~8,~10 and~12. Thus, columns~14 and~15 define the range where the central 2/3 of the datapoints are located.}
\centering
\vspace{2mm}
\begin{tabular}{rrrrrrrrrrrrrrr}
\hline
&&&&&&&&&&&&&&\\[-10pt]
$x$      & range & $N$  & $a_{mr}$ & & $b_{mr}$ & & $c_{mr}$ &  & $d_{mr}$  &  &  $e_{mr}$ & & $\Delta_-$ & $\Delta_+$ \\
\hline
$\log_{10}(L_u/{\rm L}_{\odot})$ & 3 to 10.5 & $460$  & $0.436$ & $(3.5\%)$   & $4.06$ & $(3.8\%)$   & $9.85$ & $(9.4\%)$    & $0.044$ & $(4.0\%)$  & $0.562$ & $(2.2\%)$  & $-0.0890$ &  $0.0807$ \\
$\log_{10}(L_g/{\rm L}_{\odot})$ & 3 to 11   & $460$  & $0.369$ & $(1.9\%)$   & $3.47$ & $(2.0\%)$   & $16.6$ & $(7.6\%)$    & $0.044$ & $(2.8\%)$  & $0.570$ & $(1.6\%)$  & $-0.0865$ &  $0.0733$ \\
$\log_{10}(L_r/{\rm L}_{\odot})$ & 3 to 11   & $460$  & $0.371$ & $(2.0\%)$   & $3.54$ & $(2.1\%)$   & $15.6$ & $(7.3\%)$    & $0.047$ & $(2.7\%)$  & $0.538$ & $(1.8\%)$  & $-0.0843$ &  $0.0714$ \\
$\log_{10}(L_i/{\rm L}_{\odot})$ & 7 to 11   & $393$  & $0.381$ & $(3.6\%)$   & $3.70$ & $(3.7\%)$   & $14.9$ & $(10.1\%)$   & $0.043$ & $(25.1\%)$ & $0.587$ & $(8.7\%)$  & $-0.0798$ &  $0.0743$ \\
$\log_{10}(L_z/{\rm L}_{\odot})$ & 7 to 11.5 & $347$  & $0.351$ & $(3.4\%)$   & $3.41$ & $(3.4\%)$   & $17.7$ & $(10.9\%)$   & $0.046$ & $(16.6\%)$ & $0.543$ & $(12.0\%)$ & $-0.0874$ &  $0.0762$ \\
$\log_{10}(L_U/{\rm L}_{\odot})$ & 7 to 11   & $429$  & $0.377$ & $(3.1\%)$   & $3.55$ & $(3.1\%)$   & $16.8$ & $(9.6\%)$    & $0.049$ & $(16.0\%)$ & $0.527$ & $(12.2\%)$ & $-0.0874$ &  $0.0793$ \\
$\log_{10}(L_B/{\rm L}_{\odot})$ & 3 to 11   & $460$  & $0.398$ & $(2.7\%)$   & $3.82$ & $(2.9\%)$   & $10.8$ & $(7.9\%)$    & $0.041$ & $(3.4\%)$  & $0.577$ & $(1.7\%)$  & $-0.0844$ &  $0.0739$ \\
$\log_{10}(L_V/{\rm L}_{\odot})$ & 3 to 11.5 & $460$  & $0.383$ & $(2.1\%)$   & $3.70$ & $(2.3\%)$   & $13.3$ & $(6.9\%)$    & $0.046$ & $(2.7\%)$  & $0.537$ & $(1.7\%)$  & $-0.0794$ &  $0.0697$ \\
$\log_{10}(L_I/{\rm L}_{\odot})$ & 7 to 11   & $347$  & $0.378$ & $(3.9\%)$   & $3.67$ & $(3.9\%)$   & $15.8$ & $(10.6\%)$   & $0.058$ & $(17.8\%)$ & $0.433$ & $(13.1\%)$ & $-0.0762$ &  $0.0706$ \\
$\log_{10}(M_{\rm can}/{\rm M}_{\odot})$ & 3 to 12 & $460$ & $0.679$ & $(1.4\%)$ & $7.35$ & $(1.6\%)$ & $3.92$ & $(2.4\%)$   & $0.044$ & $(0.86\%)$ & $0.544$ & $(0.54\%)$ & $-0.0119$ &  $0.0104$ \\
&&&&&&&&&&&&&&\\[-10pt]
\hline
\end{tabular}
\end{table}
\end{landscape}

The impact of the IGIMF on the mass of the stellar population of ETGs stellar mass can be parametrised as
\begin{equation}
\label{eq:Mratiofit}
\frac{M_{\rm IGIMF}}{M_{\rm can}}=\frac{a_{mr} \, x-b_{mr}}{1 -\exp[-c_{mr} (a_{mr} \, x-b_{mr})]}+d_{mr} \, x+e_{mr},
\end{equation}
where $x$ is the base-10 logarithm of a luminosity or the stellar mass under the assumption that the IMF is canonical (taken from \citealt{Dabringhausen2016a}) and $a_{mr}$, $b_{mr}$, $c_{mr}$, $d_{mr}$ and $e_{mr}$ are parameters that are obtained in least-square fit and are listed in table~(\ref{tab:Mratiofit}). 

Figures~(\ref{fig:IGIMF_LV}) and~(\ref{fig:IGIMF_Lr}) show equation~(\ref{eq:Mratiofit}) with some exemplary sets of parameters and the data to which they are fitted. The dependency of equation~(\ref{eq:Mratiofit}) is on $\log_{10} (L_V)$ in Figure~(\ref{fig:IGIMF_LV}) and on $\log_{10} (L_r)$ in Figure~(\ref{fig:IGIMF_Lr}). The $V$-band luminosities are the basis for all mass estimates in this paper, and Figure~(\ref{fig:IGIMF_LV}) thereby shows the maximum number of ETGs used in this paper, while the $r$-band is popular in studies based on the SDSS-survey.

Note that Figs.~(\ref{fig:IGIMF_LV}) and~(\ref{fig:IGIMF_Lr}) show mass ratios, not absolute masses. They are thus, more than anything else, an answer to the question how much the transition from the canonical IMF to the IGIMF with its presumably different high-mass slope changes the real mass of the galaxies relative to their masses with the canonical IMF. This property of Figs.~(\ref{fig:IGIMF_LV}) and~(\ref{fig:IGIMF_Lr}) is a consequence of the implementation of the IGIMF used here, and similar to other authors; e.g. \citet{Fontanot2017} and \citet{Yan2017}. With this implementation, the parameter quantifying the high-mass slope of the IGIMF implies that the deviation of the IGIMF from the canonical IMF is the most extreme for the most massive stars. Below and near one Solar mass on the other hand, i.e. in the mass range where the stellar population still produces a substantial optical spectrum since the stars have not evolved into remnants yet, the canonical IMF and the IGIMFs are quite similar to each other (see Sec.~\ref{sec:masses-ETGs}). Thus, the errors in luminosity made by estimating the stellar masses from a single SSP instead with a more elaborate method are expected to be negligent to a large extent. Obtaining very accurate estimates for $M_{\rm can}$ is therefore not essential to find useful values for the mass ratios, and the estimates of $M_{\rm can}$ from single SSP-models from \citet{Dabringhausen2016a} should certainly be sufficient for that.

The fact that figures~(\ref{fig:IGIMF_LV}) and~(\ref{fig:IGIMF_Lr}) show mass ratios instead of absolute values explains also why the small ETGs have only little uncertainty in the mass ratio and that the uncertainty grows with the mass of the galaxies. The small ETGs have low star formation rates which translate into high values for $\alpha_{\rm eff}$ according to the IGIMF-model adopted here. Although the values for $\alpha_{\rm eff}$ vary in the low-mass ETGs as much, or even more, with a changing mass like in the more massive ETGs, the effect on the mass ratio is small compared to the more massive ETGs. In other words, it makes much more of a difference for the actual mass of a galaxy compared to the mass it would have if it had formed with the canonical IMF, if $\alpha_{\rm eff}$ varies between 1.5 and 2 (like in a massive ETG) than when $\alpha_{\rm eff}$ varies between 3 and 3.5 (like in an low-mass ETG). Note however that the luminosity of the ETGs is not affected by the way their mass is modelled, as can be seen that each ETG is represented by a single luminosity, which is the observed one.

While the true turn-off mass in the ETGs can be replaced with $1 \, {\rm M}_{\odot}$ for the sake of dealing with the mass ratios between the canonical IMF and the IGIMF, the same is not necessarily true also for the mass-to-light ratios of individual ETGs. However, the main concern of this paper is the change of the {\it average} properties of the ETGs with their luminosity, and not precise estimates for individual ETGs. According to Fig.~(\ref{fig:SSP-comp}), the average offset due to estimating the stellar masses from a single SSP-model instead of multiple SSP-models for each ETG is about 10 per cent. This implies that also the data on mass-to-light ratios and fits to them are on average 10 per cent too high when mass estimates from single SSP-models are used for the ETGs. The impact of the IGIMF on the mass estimates is expected to be much stronger especially at high luminosities. Thus, to a good approximation, Equation~(\ref{eq:Mratiofit}) can also be seen as providing the factor by which $M/L_{pb}$-ratios coming from SSP-models with the canonical IMF have to be multiplied in order to obtain the $M/L_{pb}$-ratios according to the IGIMF model.

\begin{figure}
\centering
\includegraphics[scale=0.85]{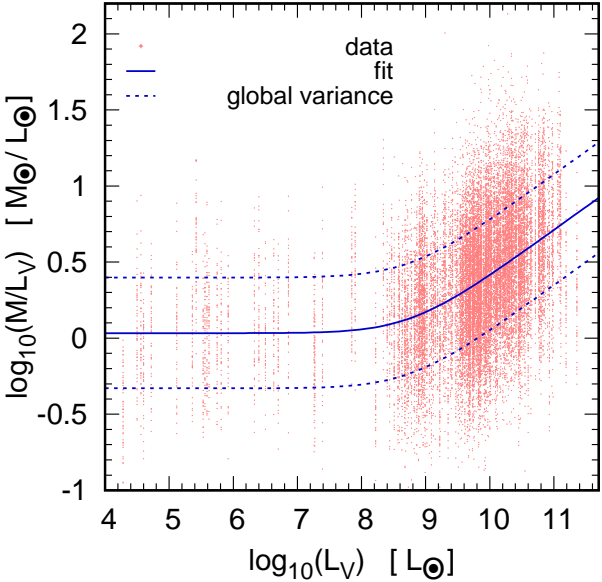}
\caption[The $M/L_V$-ratio of ETGs]{\label{fig:MLratio-LV} The mass-to-light ratio in the $V$-Band, $M/L_V$, in dependency of the $V$-band luminosity, $L_V$. Each ETG is represented 50 times with different masses that are based on the mass given in \citet{Dabringhausen2016a}, but have been modified by adding random representations of a gaussian distribution with a width of $\sigma=0.3$. The drawn line is a fit of equation~(\ref{eq:MLratiofit}) to the data. The dashed lines are same as the drawn line, except that constant values have been substracted or added to it. The range between the two dashed lines is the range in which the central 2/3 of the data lie.}
\end{figure}

\begin{figure}
\centering
\includegraphics[scale=0.85]{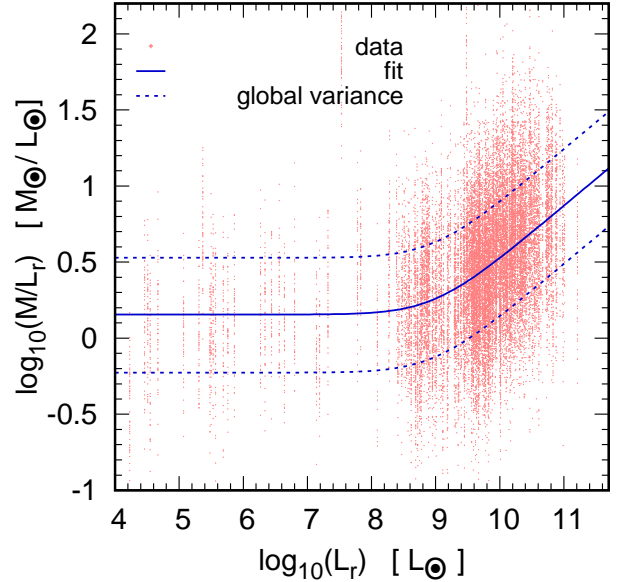}
\caption[The $M/L_r$-ratio of ETGs]{\label{fig:MLratio-Lr} As figure~(\ref{fig:MLratio-LV}), but for the $r$-band instead of the $V$-band.}
\end{figure}

The actual $M/L_{pb}$-ratios of ETGs according to the IGIMF model can also be parametrized as
\begin{equation}
\label{eq:MLratiofit}
\log_{10}\left(\frac{M}{L_{pb}}\right)=\left(\frac{a_{ML} \, x-b_{ML}}{1 -\exp[-c_{ML} (a_{ML} \, x)-b_{ML}]}+e_{ML}\right)
\end{equation}
where $x$ is the base-10 logarithm of a luminosity (taken from \citealt{Dabringhausen2016a}) and $a_{ML}$, $b_{ML}$, $c_{ML}$ and $e_{ML}$ are parameters that are obtained in least-square fits and are listed in table~(\ref{tab:MLratiofit}). Exemplary cases for the data and the fits to them are shown in Figs~(\ref{fig:MLratio-LV}) and~(\ref{fig:MLratio-Lr}).

Figures~(\ref{fig:IGIMF_LV}) to (\ref{fig:MLratio-Lr}) indicate that for low and moderate black hole masses and SFRs, the total masses of the most luminous ETGs are with the IGIMF a factor of 1.5 to 2 higher than expected with the canonical IMF. This is consistent with the difference between mass estimates for ETGs from the canonical IMF and the mass estimates for the same galaxies from gravitational lensing in \citet{Leier2016}, see their table~(3). The fits to the $M/L_V$ ratios in Fig.~(\ref{fig:MLratio-LV}) and the $M/L_r$-ratios in Fig~(\ref{fig:MLratio-Lr}) show that this increase of the mass leads to average $V$-band and $r$-band mass-to-light ratios of approximately 6 or 7 in Solar units for the most luminous ETGs. About the same values are found by \citet{LaBarbera2013} for the dynamical $M/L_r$ ratios of massive ETGs, see their figure~(21).

\begin{table*}
\caption{\label{tab:MLratiofit} Best-fitting parameters in equation~(\ref{eq:MLratiofit}) for different passbands. Column~1 specifies the variable considered in equation~(\ref{eq:MLratiofit}) and Column~2 gives the approximate range over which the function was fitted. Column~3 gives the number of galaxies used in the fit of the parameters. Columns 4,~6 and~10 list the obtained values for the parameters $a_{ML}$, $b_{ML}$ and $e_{ML}$, respectively. The numbers in brackets in columns~5,~7 and~11 are the asymptotic standard errors in per cent to the values before them. Thus, a value of more than 20 per cent shows that the asymptotic standard error is less than five $\sigma$ different from zero, provided the distribution is Gaussian. The $(*)$ in column~9 indicates that $c_{\alpha}$ in column~8 was not considered as a free parameter in the fit, but set to 10 in order to ensure convergence of the fitting routine. Column~12 lists the value that to be substracted from equation~(\ref{eq:MLratiofit}) that globally 1/6 of the datapoints are below the function defined by the parameters in columns~4,~6,~8 and~10. Column~13 lists the value that to be added to equation~(\ref{eq:MLratiofit}) that globally 1/6 of the datapoints are above the function defined by the parameters in columns~4,~6,~8 and~10. Thus, columns~12 and~13 define the range where the central 2/3 of the datapoints are located.}
\centering
\vspace{2mm}
\begin{tabular}{rrrrrrrrrrrrr}
\hline
&&&&&&&&&&&&\\[-10pt]
$x$      & range & $N$  & $a_{ML}$ & & $b_{ML}$ & & $c_{ML}$ &  &  $e_{ML}$ & & $\Delta_-$ & $\Delta_+$    \\
\hline
$\log_{10}(L_u/{\rm L}_{\odot})$ & 3 to 10.5 & $460$  & $ 0.402$ & $(1.8\%)$   & $3.26$ & $(2.3\%)$   & $10$ & $(*)$  & $ 0.204$ & $(6.1\%)$ &  $-0.382$ &  $0.420$ \\
$\log_{10}(L_g/{\rm L}_{\odot})$ & 3 to 11   & $460$  & $ 0.411$ & $(2.7\%)$   & $3.76$ & $(3.1\%)$   & $10$ & $(*)$  & $ 0.302$ & $(2.8\%)$ &  $-0.402$ &  $0.392$ \\
$\log_{10}(L_r/{\rm L}_{\odot})$ & 3 to 11   & $460$  & $ 0.353$ & $(2.4\%)$   & $3.17$ & $(2.9\%)$   & $10$ & $(*)$  & $ 0.156$ & $(5.8\%)$ &  $-0.382$ &  $0.373$ \\
$\log_{10}(L_i/{\rm L}_{\odot})$ & 7 to 11   & $393$  & $ 0.411$ & $(4.6\%)$   & $4.00$ & $(5.1\%)$   & $10$ & $(*)$  & $ 0.238$ & $(3.7\%)$ &  $-0.401$ &  $0.373$ \\
$\log_{10}(L_z/{\rm L}_{\odot})$ & 7 to 11.5 & $347$  & $ 0.376$ & $(3.5\%)$   & $3.56$ & $(4.1\%)$   & $10$ & $(*)$  & $ 0.017$ & $(98.0\%)$&  $-0.386$ &  $0.370$ \\
$\log_{10}(L_U/{\rm L}_{\odot})$ & 7 to 11   & $429$  & $ 0.417$ & $(2.4\%)$   & $3.70$ & $(3.1\%)$   & $10$ & $(*)$  & $ 0.199$ & $(8.0\%)$ &  $-0.402$ &  $0.399$ \\
$\log_{10}(L_B/{\rm L}_{\odot})$ & 3 to 11   & $460$  & $ 0.361$ & $(2.5\%)$   & $3.25$ & $(3.0\%)$   & $10$ & $(*)$  & $ 0.147$ & $(6.8\%)$ &  $-0.394$ &  $0.391$ \\
$\log_{10}(L_V/{\rm L}_{\odot})$ & 3 to 11.5 & $460$  & $ 0.304$ & $(2.2\%)$   & $2.67$ & $(2.7\%)$   & $10$ & $(*)$  & $ 0.033$ & $(28.9\%)$&  $-0.363$ &  $0.366$ \\
$\log_{10}(L_I/{\rm L}_{\odot})$ & 7 to 11   & $347$  & $ 0.420$ & $(2.9\%)$   & $3.87$ & $(3.4\%)$   & $10$ & $(*)$  & $ 0.055$ & $(22.7\%)$&  $-0.382$ &  $0.374$ \\
$\log_{10}(M_{\rm can}/{\rm M}_{\odot})$ & 3 to 12 & $460$ & $ 0.468$ & $(0.92\%)$ & $4.37$ & $(1.1\%)$ & $10$ & $(*)$ & $-0.051$ & $(9.8\%)$&  $-0.264$ &  $0.264$ \\
&&&&&&&&&&&&\\[-10pt]
\hline
\end{tabular}
\end{table*}

\section{Discussion}
\label{sec:discussion}

A popular convention for a top-heavy stellar mass function is a mass function that has more massive stars than expected for a reference mass function with a high-mass slope $\alpha_{\rm eff}<2.3$ for stars above $1 \, {\rm M}_{\odot}$ and an upper mass limit for stars of $m_{\rm max}=100 \, {\rm M}_{\odot}$. This reference mass function is implemented in widely used SSP-models like the ones by \citet{Bruzual2003} and \citet{Maraston2005}, and thereby also serves as the standard for estimating masses (or other properties) of ETGs with 'normal' stellar populations. The mass estimates thereby made correspond to the estimates for $M_{\rm can}$, and generally not to $M_{\rm IGIMF}$ (see Section~\ref{sec:data-masses}). Figures~(\ref{fig:IGIMF_LV}) and~(\ref{fig:IGIMF_Lr}) indicate that by the above standard, the stellar mass function of ETGs becomes, quite independent of the considered passband, effectively top-heavy for luminosities $L_{pb} \apprge \ 10^9 {\rm L}_{\odot}$, and top-light below this threshold value. Thus, even though common for ETGs of any size, using SSP-models for stellar populations with the canonical IMF directly for mass estimates of ETGs leads to slightly overestimating the mass of ETGs with luminosities $L_{pb} \apprle \ 10^9 {\rm L}_{\odot}$, and to underestimating, possibly quite severely, the masses of the most massive ETGs.

The coincidence of $M_{\rm IGIMF}$ with $M_{\rm can}$ at $L_{pb} \approx \ 10^9 {\rm L}_{\odot}$ indicates that ETGs with a luminosity $L_{pb} \apprge \ 10^9 {\rm L}_{\odot}$ must have hosted quite a number of star clusters with a top-heavy IMF. The reason is that according to the IGIMF-model, all galaxies also form low-mass star clusters in which massive stars do not form, and if the galaxy as a whole is characterized with $\alpha_{\rm eff}=2.3$, it must therefore also have formed massive star clusters with a top-heavy IMF (i.e. $\alpha_3<2.3$ in equation~\ref{eq:IMF}), in order to compensate the lack of high-mass stars in the low-mass clusters. In other words, in a galaxy where all star clusters have formed with the canonical IMF, the IGIMF must be steeper. In the Milky Way, $\alpha_{\rm eff}>2.3$ is indeed the case, which is consistent with the claimed ubiquity of the canonical IMF in star clusters in the Milky Way \citep{Kroupa2001}. Thus, in summary, and using the definitions given at the end of Section~(\ref{sec:simplification}), the IGIMF of ETGs with $L_{pb} \apprge 10^9 \ {\rm L}_{\odot}$ is top-heavy, and the IGIMF of ETGs with $L_{pb} \apprle 10^9 \ {\rm L}_{\odot}$, and the Milky Way, is top-light.

ETGs are dominated by old stellar populations, which means that massive stars in the mass-range where the IGIMF varies with the SFR have already evolved into remnants. This implies that massive ETGs with their high SFRs are expected to have higher $M/L_{pb}$-ratios than expected from SSP-models with the canonical IMF. Observational data on ETGs confirm this. The detected disagreement between predictions for the masses from SSP models with the canonical IMF and estimates for the masses from internal dynamics is not as dramatic as in spiral galaxies or dwarf spheroidal galaxies, but well documented. As one possibility to explain this, non-baryonic dark matter was considered (e.g. \citealt{Cappellari2006,Tortora2009}). Milgromian dynamics (MOND, \citealt{Milgrom1983}), on the other hand, does not influence the internal dynamics of ETGs enough to produce the observed discrepancy, as long as the assumption of a SSP with $\alpha_3=2.3$ and $m_{\rm max}= 100 \ {\rm M}_{\odot}$ is maintained \citep{Tortora2014,Dabringhausen2016b}. As an alternative explanation for the high $M/L_{pb}$-ratios of ETGs, variations of the IMF, or better variations of the IGIMF\footnote{Note that many authors cited in this section discuss variations of the IMF, even though they discuss in fact the stellar mass functions of whole galaxies instead of individual star clusters within those galaxies. Thus, following the nomenclature used in this paper, they rather consider IGIMFs instead of IMFs. This can lead to some confusion when the terms 'top-heavy' and 'top-light' are used, as the example of the Milky Way may show: The Milky Way forms star clusters with a high-mass IMF-slope of $\alpha_3=2.3$, i.e. star clusters whose IMFs would not be perceived as top-light, but due $m_{\rm max} < m_{\rm max *}$ in low-mass star clusters, the IGIMF of the Milky way is nevertheless top-light, i.e. $\alpha_{\rm eff}>2.3$. In this paper, we therefore try to use term 'IGIMF' consistently if the overall stellar population of a galaxy is considered, even if the term 'IMF' is used in the cited paper.}, have been suggested, and the required amount of matter that is unaccounted for can indeed be provided by variations which seem reasonable in their magnitude, especially in combination with MOND \citep{Tortora2014,Dabringhausen2016b}.

The $M/L_{pb}$-ratios as such do not put strong constraints on the type of IGIMF-variation that cause them. In old stellar populations as they are typical for ETGs, high $M/L_{pb}$-ratios can either be the consequence of an IGIMF that is skewed to low-mass stars (i.e. bottom-heavy), which implies large population of faint low-mass stars, or the consequence of top-heavy IGIMF which implies a large population of essentially non-luminous remnants of massive stars \citep{Cappellari2012}, or both \citep{Jerabkova2018}. A bottom-heavy IGIMF in massive ETGs has for instance been proposed in \citet{Samurovic2010} and \citet{Tortora2014} based on their internal dynamics, but also in \citet{vanDokkum2010} based on a different indicator, namely the strength of some absorption lines which are sensitive to the presence of low-mass stars. The IGIMF-model on the other hand predicts that the effective stellar mass function in massive ETGs would effectively become top-heavy because they formed most of their stars at high SFRs, as has been quantified e.g. in \citet{Weidner2011}, \citet{Weidner2013b} and \citet{Fontanot2017}. \citet{Jerabkova2018} suggests that certain galaxies can have IGIMFs that are top-heavy {\it and} bottom-heavy at the same time, which is possible since top-heaviness and bottom-heaviness address different parts of the mass spectrum of stars.

There is indeed observational evidence that star-burst systems, as massive ETGs certainly were in the past (see, e.g. figure~10 in \citealt{Thomas2005}), had a top-heavy IGIMF. For instance, \citet{vanDokkum2008} studied the luminosity evolution and the colour evolution of massive ETGs and concluded that the combination of the two implies that their IGIMFs were likely to be top-heavy in the past. \citet{Gunawardhana2011} correlated H$\alpha$ luminosities (which are an indicator for the star formation rate) with optical colours (which are an indicator for the composition of the stellar population), and found on this basis that the IGIMF in ETGs is dependent on the SFR, such that ETGs with $\log_{10}(SFR/{\rm M}_{\odot}\, yr^{-1})<0$ tend to have top-light IMFs and ETGs with $\log_{10}(SFR/{\rm M}_{\odot}\, yr^{-1})>1$ tend to have top-heavy IMFs (see their figure~5). Most ETGs with $0<\log_{10}(SFR/{\rm M}_{\odot}\, yr^{-1})<1$ have IGIMFs with $\alpha_{\rm eff} \approx 2.3$ according to \citet{Gunawardhana2011}, which is remarkably well consistent with the quantification of the IGIMF used in the present paper (see Figure~\ref{fig:SFR-alpha_eff}). \citet{Romano2017} use a different approach by studiying the abundances of CNO and linking them to different evolving stars, but again arrive at the conclusion that the abundances in starbursting systems suggest a top-heavy IGIMF in them. Finally, it turns out that the $M/L_{pb}$-ratios predicted by the IGIMF-model for stellar populations of ETGs are indeed very well consistent with the masses implied by their internal dynamics (Dabringhausen et al., in preparation).

A possible way to reconcile the evidence for bottom-heavy IGIMFs in massive ETGs coming from spectral lines with the evidence for top-heavy IGIMFs in the same systems is a variation of the IGIMF over time. Star formation in the massive ETGs may have started with a top-heavy IMF, but continued with a bottom-heavy IGIMF later on. The reason for this may be that the rapid formation of massive stars early in the life of the massive ETGs would certainly have increased the metallicity of the star-forming interstellar medium quickly, and perhaps also its turbulence \citep{Weidner2013a}. It has indeed been argued that the shape of the IMF, and thus also the shape of the IGIMF, is skewed towards lower stellar masses for higher metallicities (see e.g. \citealt{Marks2012a}). The IGIMF is according to this notion primarily a metallicity effect, and the SFR is only a proxy that measures the expected strength of this effect. 

In a more recent work, \citet{Jerabkova2018} considered that the IGIMF does not depend on only one parameter, but two parameters, namely the metallicity besides the SFR. The metallicity dependency is taken in \citet{Jerabkova2018} from \citet{Marks2012b}, i.e. it is the metallicity dependency that is neglegted here for simplicity over the much stronger SFR-dependency. However, implementing the metallicity as well does indeed to IGIMFs that are both top-heavy and bottom-heavy at the same time in galaxies that have high SFRs and high metallicities; i.e. the massive ETGs. Thus, it seems that considering the metallicity as well can solve the controversy whether the massive ETGs have a bottom-heavy or a top-heavy IGIMF by letting them have both, depending on their metallicity.

\section[Conclusion]{Summary and Conclusion}
\label{sec:conclusion}

Observations supported for quite some time that the stellar initial mass function (IMF) in star clusters is at least in the Local Group invariant \citep{Kroupa2001}, even though theoretical considerations make a variation of the IMF with the properties of the environment appear likely (e.g. \citealt{Murray1996,Larson1998}). More recent research has finally provided evidence for the notion that the IMF in a star cluster changes with its properties. This is for low-mass clusters a dependence of the maximum mass of a star that can form in a cluster on the cluster mass \citep{Weidner2005,Weidner2006,Weidner2010}, and for high-mass clusters a dependence of the IMF-slope on the cluster mass \citep{Dabringhausen2009,Dabringhausen2012,Marks2012a}. These variations of the IMF also affect the overall stellar populations in galaxies, since the mass distribution of the star clusters born in a galaxy depend on its star formation rate (SFR) in the galaxy \citep{Weidner2004b}. Thus, galaxies with different global SFRs have different populations of star clusters and thus different overall stellar populations. This is quantified in the IGIMF-model (e.g. \citealt{Weidner2005,Weidner2011,Weidner2013b,Fontanot2017,Yan2017}).

The existing quantifications of the IGIMF mostly express the shape of the stellar mass spectrum in dependency of the SFR. While the variation of the IGIMF from galaxy to galaxy can indeed be expressed with the SFR being the sole free parameter, the practical disadvantage of this approach is that the SFR is one of the parameters of a galaxy which is more difficult to estimate. That is to say that the overall properties of ETGs are dominated by old stellar populations, which formed when the SFRs were quite different from what they are today. However, the most likely formation timescales of ETGs can estimated, which, together with an estimate of the masses of the ETGs, can be turned into an estimates for their characteristic SFRs at the time when most of their stars formed. Fortunately, these characteristic past SFRs and the corresponding IGIMF can be linked to more intuitive, more accessible and more widely used quantities like the present-day luminosities in various passbands. This is done in this paper, using the data on the luminosities and masses of ETGs in \citet{Dabringhausen2016a}. Moreover, parametrisations of the shape of the IGIMF and its consequences on the luminosities in various passbands are provided here.

In particular, the variation of the IGIMF, including the upper mass limit, is parametrised as a function of the SFR, using the data found in \citet{Fontanot2017}. They provide data only for a grid of specific SFRs, but this grid is replaced here by a continuous formulation. The parametrisation of the IGIMF introduced here also includes a simplification of the shape of the IGIMF proposed in \citet{Fontanot2017} by reducing the number of parameters that determine the shape of the IGIMF for stellar masses $m\apprge 1 \ {\rm M}_{\odot}$ from five (two high-mass slopes $\alpha_3$ and $\alpha_4$, a transition mass where the slope changes from $\alpha_3$ to $\alpha_4$, a transition mass near $1 \ {\rm M}_{\odot}$ where the IGIMF-slope at lower stellar masses changes to $\alpha_3$, and the upper mass limit for stars that can form, $m_{\rm max}$) to two. These are $m_{\rm max}$ and a single effective high-mass IGIMF slope $\alpha_{\rm eff}$. The transition from the slope at lower stellar masses to $\alpha_{\rm eff}$ is fixed to exactly $1 \, {\rm M}_{\odot}$. The resulting parametrisation of the IGIMF should still be accurate enough for practical purposes, since the properties of the stellar populations of galaxies do not only depend on the precise shape of the IGIMF and how it varies with the SFR, but also on the star formation history of the galaxies, i.e. how the SFR of the galaxies changes with time. However, reconstructing the SFH of a galaxy beyond a very general picture is very challenging \citep{Maschberger2007}.  

Moreover, the variation of $\alpha_{\rm eff}$ and its consequences for the masses of stellar populations of the ETGs in \citet{Dabringhausen2016a} are estimated here. For this, the correlation between the stellar masses and the timescales for the formation of these galaxies \citet{Thomas2005,Recchi2009} is used. For the estimates of $\alpha_{\rm eff}$ and the masses of the stellar populations of the ETGs, also interpolation functions are fitted, which allow to make simple estimates of these parameters for them, based on their luminosities.

It turns out from these estimates that $\alpha_{\rm eff}=2.3$ for ETGs with a luminosity of about $10^9 \ {\rm L}_{\odot}$ for any of the passbands considered here. This means that at this luminosity, the IGIMF for the ETGs is identical to the IMF for star clusters as formulated in equation~(\ref{eq:IMF}), with the maximum stellar mass, $m_{\rm max}$, corresponding to the physical upper mass limit for stars, $m_{\rm max*}$. In consequence, the common practise of estimating the properties of ETGs (such as total stellar mass and colours) from SSP-models with the canonical IMF (like the ones by \citealt{Bruzual2003} or \citealt{Maraston2005}) leads to the correct results only at this luminosity.

For ETGs with luminosities $L_{pb} \apprle 10^9$, the IGIMF model implies $\alpha_{\rm eff}>2.3$. In consequence, these ETGs have not formed as many massive stars as it would be expected based on a SSP-model with the canonical high-mass IMF slope $\alpha_3=2.3$ and an upper stellar mass limit of $m_{\rm max}=m_{\rm max*} \approx 150 \ {\rm M}_{\odot}$. On the other hand, stars with masses $m \approx 150 \ {\rm M}_{\odot}$ may also form in ETGs with luminosities $10^8 \ {\rm L}_{\odot} \apprle L_{pb} \apprle 10^9 \ {\rm L}_{\odot}$. The reason for $\alpha_{\rm eff}>2.3$ in these galaxies is that essentially all their stars formed in low-mass star clusters with $m_{\rm max}<m_{\rm max*}$, since the low SFR suppressed the formation of high-mass star clusters with $m_{\rm max}=m_{\rm max*}$ and $\alpha_3<2.3$. Thus, the IGIMFs of ETGs with luminosities $L_{pb} \apprle 10^9$ can be considered top-light in comparison to the canonical IMF. The contrary is true for ETGs with with luminosities $L_{pb} \apprge 10^9$. For them, the IGIMF model implies $\alpha_{\rm eff}<2.3$ and an upper mass limit of $m_{\rm max}=m_{\rm max*} \approx 150 \ {\rm M}_{\odot}$. Thus, the IGIMF in such ETGs can be considered top-heavy in comparison to the canonical IMF.

Since most of the stars in ETGs are at least a few Gyrs old, the massive stars in ETGs have already evolved into remnants and thereby only add to the masses of the ETGs, but not to their luminosities. In consequence, estimates for the masses of the stellar populations of ETGs with $L_{pb} \apprle 10^9 \ {\rm L}_{\odot}$ are too high by a factor up to approximately 1.5, if they are done with the above mentioned SSP-models without correcting for effect of the IGIMF. The missing mass problem, i.e. the discrepancy between mass estimates for ETGs based on their internal dynamics and mass estimates based on the observed baryonic matter in them (e.g. \citealt{Mateo1998,Wolf2010,Dabringhausen2016b,Lelli2017}) is thereby worsened, even though not by much. Provided the ETGs are in dynamical equilibrium (which has been questioned by many authors, e.g. \citealt{Kroupa1997,Metz2007,McGaugh2010,Casas2012,Dominguez2016}), the internal dynamics of very faint ETGs suggest masses up to a thousand times larger larger than what can be explained with their stellar populations, and overestimating the mass of the stellar population by a factor of approximatey 1.5 is a minor error in comparison.

For ETGs with luminosities $L_{pb} \apprge 10^9$, the IGIMF model implies $\alpha_{\rm eff}<2.3$, since the top-heavy IMF in massive star clusters overcompensates the lack of massive stars in low-mass star clusters in these galaxies. By how much the masses of the stellar populations (including remnants) of such massive ETGs exceed the expectations based on SSP models with $\alpha_3=2.3$ and an upper stellar mass limit of $m_{\rm max}=m_{\rm max*} \approx 150 \ {\rm M}_{\odot}$ depends quite significantly on the assumptions regarding the SFR and the mass of the remnants. The reasons are that at low values of $\alpha_{\rm eff}$, slight changes of $\alpha_{\rm eff}$ have a rather strong impact on the number of remnants in a galaxy, and the assumptions regarding the mass of the remnants become quite important if the remnants are numerous. The best assumptions for the effective SFR of the ETGs and the mass of stellar-mass black holes (i.e. the remnants of the most massive stars) are however rather uncertain parameters. A constant global SFR for a galaxy estimated from its mass and a timescale for star formation can only be an approximation to its real SFR over time for several reasons. First, the SFR will not be constant in a real galaxy, but is expected to vary over time (see e.g. figure~1 in \citealt{delaRosa2011}) Moreover, even at a given time, the local SFR at the dense centre of a galaxy is likely to be higher than at its flimsier outskirts, which is supported by the gradients of the stellar mass function found by \citet{MartinNavarro2015} and \citet{vanDokkum2017} in massive ETGs. The departure from the canonical IMF is the strongest in the central parts of these ETGs. Finally, the evolution of the SFR of a galaxy over time can also be influenced by interactions and mergers with other galaxies, as is illustrated by the starbursts in merging systems like the Antennae Galaxies (NGC4038 and NGC4039). Nevertheless a dependency of the IGIMF on the global time-averaged SFR has been established \citep{Weidner2004b,Weidner2005}, even though the detailed physical reasons for this dependency are not well understood \citep{Weidner2013a}. The SFRs and IGIMFs of individual ETGs may deviate significantly from the correlations established for populations of such objects. An uncertainty for the mass of the remants comes from the fact it depends quite strongly on the metallicity of the progenitor stars, which changes significantly over time by the enrichment of the interstellar medium by evolving stars. Thus, just like a global time-averaged SFR can only be an approximation to its actual SFR over time and space, assuming that the mass of stellar remnants is given as a constant fraction of the mass of the progenitor stars can only be an approximation to the real masses of the remnants.

In this paper, the problem with the variability and uncertainty of the SFR is solved by calculating the time scale for star formation with equation~(\ref{eq:deltatMdyn}, i.e. equation~5 in \citealt{Thomas2005}) for an ETG, and then dividing the total stellar mass of the ETG by the result to get an estimate for its SFR. The problem with the mass dependency of stellar remnants with metallicity is solved with equation~(\ref{blackhole}), where the metallicity is taken as the present-day values of the ETGs. These approximations are admittedly pretty basic, but the resulting masses of the ETGs with IGIMFs are remarkably well consistent with earlier results based on the dynamics or gravitational lensing (e.g. \citealt{LaBarbera2013,Leier2016}). For the most massive ETGs, they are up to a factor of 2 higher than mass estimates with the canonical IMF with $m_{\rm max*} \approx 150 \ {\rm M}_{\odot}$.

Thus accounting for the effect of the IGIMF decreases the mass estimates for low-mass ETGs and increases them for high-mass ETGs in comparison to the usual estimates based on SSPs with the canonical IMF. In consequence, the missing mass problem in low-mass ETGs is somewhat aggravated, even though this may seem negligible in practise. A missing mass problem also exists in high-mass ETGs, even though to a much lower extent than in low-mass ETGs \citep{Dabringhausen2016b}. In these galaxies, accounting for the IGIMF alleviates the missing mass problem and may even solve it altogether without the assumption of additional non-baryonic matter or non-standard gravity. This will be discussed in more detail in a forthcoming paper. 

\section*{Acknowlegdements}

I thank Pavel Kroupa and Tereza Je{\v r}{\'a}bkov{\'a} for insightful discussions on the IGIMF, that led to the writing of this paper. I also thank the reviewer for the careful report on this work.

\bibliographystyle{mn2e}
\bibliography{ms}

\label{lastpage}

\end{document}